\documentclass[aps,prb,superscriptaddress,runinaddress,floatfix]{revtex4}
\usepackage{eurosym,amsfonts,amssymb,amsmath,bm,times}
\usepackage{graphicx}
\usepackage{dcolumn}
\usepackage{hyperref}
\usepackage[letterpaper,left=0.7in,right=0.7in,top=0.75in,bottom=0.55in]{geometry}

\newcommand{\func}[1]{\operatorname{#1}}
\hypersetup{colorlinks=true,linkcolor=blue,citecolor=blue,urlcolor=blue}

\begin{document}

\title{doFORC tool for calculating first-order reversal curve diagrams of
noisy scattered data}
\author{Dorin Cimpoesu}
\email{cdorin@uaic.ro}
\author{Ioan Dumitru}
\author{Alexandru Stancu}
\affiliation{Department of Physics, Alexandru Ioan Cuza University of Iasi, Iasi 700506,
Romania}

\begin{abstract}
First-order reversal curves (FORC) diagram method is one of the most
successful characterization techniques used to characterize complex
hysteretic phenomena not only in magnetism, but also in other areas of
science like in ferroelectricity, geology, archeology, light-induced and
pressure hysteresis in spin-transition materials, etc. Because the
definition of the FORC diagram involves a second-order derivative, the main
problem in their numerical calculation is that the derivative of a function
for which only discrete noise-contaminated data values are available
magnifies the noise that is inevitably present in measurements. In this
paper we present doFORC tool for calculating FORC diagrams of noise
scattered data. It can provide both a smooth approximation of the measured
magnetization and all its partial derivatives. doFORC is a free, portable
application working on various operating systems, with an easy to use
graphical interface, with four regression methods implemented to obtain a
smooth approximation of the data which may then be differentiated to obtain
approximations for derivatives. In order to perform the diagnostics and
goodness of fit doFORC computes residuals to characterize the difference
between observed and predicted values, generalized cross-validation to
measure the predictive performance, two information criteria to quantify the
information that is lost by using an approximate model, and three degrees of
freedom to compare different amounts of smoothing being performed by
different smoothing methods. Based on these doFORC can perform automatic
smoothing parameter selection.
\end{abstract}

\maketitle

\section{Introduction}

The experimental technique based on the systematic measurement of a special
category of minor loops covering the interior of the major hysteresis loop,
known as the first-order reversal curves (FORC),\cite{Mayergoyz JAP 1985,
Pike JAP 1999} is now becoming the standard characterization tool in
magnetism and in other areas of research like the ferroelectricity,\cite%
{Stancu APL 2003} geology,\cite{Roberts RG 2014} archeology,\cite{Matau JAS
2013} spin-transition thermal,\cite{Enachescu PB 2004} pressure,\cite{Tanasa
PRB 2005} light-induced hysteresis,\cite{Enachescu PRB 2005} etc. This
astonishing expansion is associated with an intense effort to improve both
the fundamental basis of the method and the computational tools accompanying
the practical use of the FORC diagrams which can be obtained from the
experimental data.

The original idea of the FORC diagram method was developed by Mayergoyz\cite%
{Mayergoyz JAP 1985} as a tool to identify the Preisach distribution in
systems correctly described by the Classical Preisach Model (CPM).\cite%
{Preisach 1935} However, in order to provide a direct way to find the
Preisach distribution associated to a given sample, Mayergoyz have shown
that the system should obey two properties: wiping-out (perfect closure of
all the minor loops and recovery of the same microscopic state after such a
loop) and congruency of the closed minor loops between the same field
limits. These two conditions proved to be too strong for the magnetic
systems and consequently the conclusion was that no real ferromagnet could
be perfectly described by a CPM with a Preisach distribution found with the
FORC technique.

Pike and collaborators\cite{Pike JAP 1999} have proposed as o solution of
this dilemma the use of the method as a purely experimental one with no link
to the Preisach model. The mathematical treatment of the experimental data
described by Mayergoyz was used to find a \textquotedblleft FORC
distribution\textquotedblright\ and not the Preisach distribution. The
concept of the \textquotedblleft magnetic fingerprinting\textquotedblright\
of samples using the FORC diagrams (contour plots of the FORC distributions)
was introduced.\cite{Katzgraber PB 2004} However, further analysis has shown
the ability of modified versions of the Preisach model to reproduce the main
features observed on the experimental FORC diagrams.\cite{Stancu JAP 2003}
This has opened again the gate towards the use of the FORC diagrams as a
quantitative tool and not merely as a qualitative one.

The conjointly use of several experimental and modeling techniques is the
current state-of-the-art in the quantitative FORC technique which already
has offered profound physical insight in the study of many magnetic systems
like the magnetic nanowire arrays,\cite{Dobrota JAP 2013, Almasi-Kashi PB
2014, Dumas PRB 2014, Gilbert SR 2014, Nica PB 2015} magnetic wires,\cite%
{Cimpoesu JAP 2016} antidote lattices,\cite{Grafe PRB 2016} magnetic bilayer
ribbons,\cite{Rivas APL 2015} nanostrips,\cite{Ognev N 2017} hard/soft
bilayer magnetic antidots,\cite{Beron NRL 2016} magnetic multilayer systems,%
\cite{Markou JMMM 2013, Davies APL 2013} and even in molecular magnets at
very low temperatures.\cite{Beron APL 2013} Many other such valuable studies
could be mentioned in the last few years alone.

As any new characterization technique, when it offers quantitative
information concerning the analyzed samples, there is a great responsibility
related to indeed know how all the practical steps involved in the
methodology are used and their possible influence in the quality of the
results. This could be simply seen as a comprehensive error analysis. Due to
the complexity of the experimental and/or numerical data (produced as a
result of various models) we put in this article emphasis on the critical
points and on the diversity of mathematical tools available to solve
properly and with the highest degree of confidence this remarkable new
technique developed in ferromagnetism, but used, as we have mentioned before
in many other studies involving hysteretic physical processes.

In this article at the beginning we briefly describe the FORC technique
itself and the few numerical packages publicly available for the data
treatment. Since in many cases the data are affected by noise the next
sections are intended to be an expository description of some of the basic
concepts related to numerical differentiation of noisy data, smoothing by
local regression, diagnostics and goodness of fit, and automatic smoothing
parameter selection. These concepts are necessary to fully understand the
capabilities of our comprehensive doFORC tool for calculating FORC diagrams.
We note, however, that we also provide a minimalist interface of doFORC
involving only the mandatory parameters that do not require understanding of
all the concepts presented in these sections. After we introduce the four
nonparametric regression methods used, we describe in detail all the
features of the doFORC. As the development of test problems is mandatory for
the validation of any numerical method, the last section of this paper is
dedicated to the test problems used to validate the algorithms used by
doFORC. We provide a dedicated graphical interface incorporating appropriate
testing tools available to users.

\section{Tools for FORC processing}

A FORC is obtained by saturating the sample under study in a positive
magnetic field $h_{\mathrm{sat}}$, decreasing the field to the reversal
field $h_{\mathrm{reversal}}$, and then sweeping the field back to $h_{%
\mathrm{sat}}$. The FORC is defined as the magnetization curve that results
when the applied field $h_{\mathrm{applied}}$ is increased from $h_{\mathrm{%
reversal}}$ to $h_{\mathrm{sat}}$, and it is a function of $h_{\mathrm{%
applied}}$ and $h_{\mathrm{reversal}}$. This process is repeated for many
values of $h_{\mathrm{reversal}}$, so that the reversal points cover the
descending branch of the major hysteresis loop (MHL), while the
corresponding FORCs cover the hysteretic surface of MHL. The FORC diagram is
defined as the mixed second derivative of the set of FORCs with respect to $%
h_{\mathrm{applied}}$ and $h_{\mathrm{reversal}}$, divided by $2$ and taken
with negative sign.\cite{Pike JAP 1999} Before each FORC measurement
saturation needs to be achieved in order to erase completely the previous
magnetic history.

If instead the sample is saturated in a negative field and accordingly the
reversal point cover the ascending branch of MHL, then the FORC diagram must
be represented as a function of \ $\left( h_{\mathrm{reversal}},h_{\mathrm{%
applied}}\right) $ in order to obtain the FORC distribution. This
distribution should be identical to the one obtained from the descending
branch of the hysteresis loop in the case of systems correctly described by
the CPM. For such systems the FORC and Preisach distributions are identical.

The main problem in the numerical calculation of the FORC diagrams is that
the second-order derivative of a function for which only discrete
noise-contaminated data values are available magnifies the noise that is
inevitably present in the magnetization measurements. Although the noise may
not particularly be evident in the original data, it is more noticeable in
the derivative. If we use conventional finite-difference approximations to
obtain the derivatives'\ values, then arbitrarily small relative
perturbations in the data can lead to arbitrarily large variations of the
results. If the available data are scattered (irregular, non-uniform
distributed) then the problem becomes even more complex.

Assuming that the data points are approximately equally spaced within the $%
\left( h_{\mathrm{applied}},h_{\mathrm{reversal}}\right) $ coordinate
system, Pike et al.\cite{Pike JAP 1999} defined a local square grid of $%
\left( 2\mathrm{SF}+1\right) ^{2}$ points surrounding a given point and the
magnetization is fitted using the least squares method with a local
second-order polynomial function. The value of the FORC distribution at the
given point is then equal to polynomial's mixed second derivative. The size
of the square grid is given by the smoothing factor SF, determining how much
of the data is used to fit each local polynomial. The entire measured
magnetization surface is fitted in a moving way, repeating the above process
for all grid points. The smoothing process is local because each smoothed
value is determined by neighboring data points defined within the local
square grid. The degree of smoothing increases with the value of SF: a large 
$\mathrm{SF}$ means that each point is more interrelated to neighboring
points, producing a smoother surface, but may result in oversmoothing
missing important features in the data, while a small $\mathrm{SF}$ may
result in a fit that is too noisy. However, the selection of an appropriate
smoothing factor in Ref. \onlinecite{Pike JAP
1999} is qualitative, being indicated that set at 2 or 3 should be
sufficient for most data sets, but a maximum of $\mathrm{SF}=5$ could be
employed for noisy curves. One drawback of the method is that no assessment
is made of the extent to which the smoothed data deviates from the measured
points.

In Ref. \onlinecite{Heslop JMMM 2005} Savitzky-Golay method is adapted to
two-dimensional case. One limitation of Savitzky-Golay approach is the
assumption that data are equally spaced. Otherwise, it is necessary to
preprocess the data using a two-dimensional interpolation to calculate the
magnetization values at points on a regular field grid. Signal-to-noise
ratio SNR for the smoothed data provides a direct measure of the deviation
from the measured data. Although SNR provides a quantitative measure of the
amount of smoothing that has taken place it gives no indication if an
appropriate value of smoothing factor was selected. The appropriate
smoothing factor is determined by performing spatial autocorrelation on the
fitting residuals.

FORCit software\cite{FORCit 2007, FORCit} is a combination of Unix shell
scripts, FORTRAN 77 subroutines, and Generic Mapping Tool (GMT) commands,
all of which are available free of charge. FORCit is executed from a
standard Unix terminal window and generates two- and three-dimensional
graphics that display the FORC distribution in a variety of coordinate
spaces as well as other related graphics. These results are saved as graphic
PostScript files, along with a Unix shell script containing information
about parameters that were used, and a GMT grid file containing the FORC
distribution as a function of coercivity and interaction. In order to remove
the noise inherent in the data FORCit first filters the gridded data using a
Gaussian filter that is part of GMT and then takes the derivatives using a
function of GMT based on central finite differences. The filter size can be
adjusted by the user within a parameter file. FORCit was written
specifically to read the output from the MicroMag instruments produced by
Princeton Measurements Corporation (PMC).

The PMC output files contain multiple header information lines, optionally
followed by the data containing the hysteresis loop, and then by the data
containing each FORC, with the first line from a FORC giving the reversal
point. Each FORC is preceded by the drift field measurement and a blank
line, and each FORC is followed by a blank line. The first FORC must contain
a single point.

FORCinel software\cite{FORCinel 2008, FORCinel} uses the LOESS smoothing
technique,\cite{Cleveland 1979, Cleveland 1988} locally fitting a
second-order polynomial function to the measured magnetization with the
weighted least squares method. The fitting is carried out over a region of
arbitrary shape containing a user-defined number of nearest-neighbor data
points, rather than a square grid of points. In order to find the optimum
smoothing factor the standard deviation of the residual $\sigma _{r}$
between the measured and the smoothed magnetization as a function of
smoothing factor is computed and the optimal smoothing corresponds to a
minimum of $\sigma _{r}$'s first derivative. The FORCinel software is freely
available, but it runs inside the commercial software Igor Pro by
Wavemetrics. Starting with version v.3.0, beside PMC file format, the
FORCinel can process also files similar with the PMC format, but without the
drift field measurements.

The VARIFORC algorithm\cite{Egli 2013} allows variable smoothing of
different parts of a FORC distribution with contrasting SNRs, and it is
implemented in FORCinel.

xFORC software\cite{xFORC} performs a non-weighted second order polynomial
regression for each data point. xFORC is freely available as an executable
file which requires LabVIEW run-time engine installed.

FORC+\cite{Abugri JAP 2018} estimates the mixed derivative of the
magnetization using finite differences without the use of smoothing in order
to obtain the sharp structures in the FORC distribution. Instead FORC+
display the raw data in such a color scheme that the noise appears grey from
a distance for the human eye.

In this paper we present the doFORC tool\cite{doFORC} for calculating FORC
diagrams of noise scattered data. It can also provide both a smooth
approximation of the measured magnetization and all its partial derivatives.
doFORC is a portable (standalone) application working on various operating
systems, and it is made freely available to the scientific community. A
portable application does not use an installer and does not modify the
existing operating system and its configuration. doFORC save the processed
data into new files whose names are obtained from the user input file.
Optionally doFORC can save configuration files containing the user's
parameters for further utilization.

A user easy to use graphical user interface (GUI) has been created
encapsulating the implemented algorithms. The interface allows users to
import the input data, set the fitting parameters, to graphical represent
(2D, 3D, and projection) both input and output data, and to export the
graphs to several image file formats. Input data may have various formats,
including the PMC MicroMag format. Output data can be provided at the input
points, on a regular grid, or at the user defined points. Even if the
program is mainly dedicated to FORC diagrams computation, it is more general
and can smooth and approximate the derivatives of a general set of
arbitrarily distributed two-dimensional points. A comprehensive description
can be found at Ref. \onlinecite{doFORC}.

The doFORC software implements four nonparametric methods for estimating
regression surface and the corresponding partial derivatives: LOESS\cite%
{Cleveland 1979, Cleveland 1988} and three modified Shepard methods\cite%
{Renka 1988 Q, Renka 1999 C, Renka 1999 T} (quadratic polynomial, cubic
polynomial, and cosine series). The motivation is to enable researchers to
experiment with different algorithms using their data and select one (or
more) that is best suited to their needs. In all four methods the problem is
addressed as first obtaining a smooth approximation of the data which may
then be differentiated to obtain approximations to the desired derivatives.
As the original modified Shepard methods were intended for interpolation we
have generalized them to smoothing methods to be able to deal with
inaccurate or noisy data. All algorithms described and GUI are implemented
in Fortran. The graphical interface is realized using the DISLIN library.%
\cite{DISLIN} The fact that both the algorithms and the interface are
implemented in Fortran allows a direct interaction between them, without the
need to use temporary working files saved on the computer.

The smoothing parameter and the local approximant function control the
amount of smoothing being performed. When smoothing is accomplished a
question to be asked is how should the smoothing parameter be chosen, how
can the performance of a smoother with a given smoothing parameter be
evaluated, how well does smooth function estimate the true function?
Choosing the smoothing parameter is a tradeoff between small bias and small
variance, but more precise characterizations are needed to derive and study
selection procedures. Another more complex and profound issue is in
comparing fits from different smoothers.

Helpful in selecting the smoothing parameter is the visual analysis of the
smoothed data for several smoothing parameters. Nevertheless, this is a
subjective method and consequently objective methods may be preferred in
order to produce an automatic smoothing selection or for consistency of
results among different investigators.

Ideal would be fully automated methods that would be able to provide the
best fit for a given data set. However, this goal is unachievable because
the best fit depends not only on the data set, but also on the questions of
interest. Nonetheless, what statistics can provide are tools to help guide
the choice of smoothing parameters, to help decide which features are real
and which are random.

In order to perform the diagnostics and goodness of fit doFORC compute the
residuals to characterize the difference between the actual observed value
and the predicted value, generalized cross-validation to measure the
predictive performance of the model, two information criteria to quantify
the information that is lost by using an approximate model on the available
data, and three degrees of freedom to compare different amounts of smoothing
being performed by different smoothing methods. Based on these criteria
doFORC can perform automatic smoothing parameter selection.

doFORC can also perform iterative reweighting to provide robust fitting when
there are outliers in the data (i.e., observation points that are distant
from other observations, having a large residual; graphically the outliers
are far from the pattern described by the other points).

To study smoothers, it is of paramount importance to use, in addition to
real data, made-up data where the true model is known. Obviously, in the
usual application no representation of the true underlying function or
surface is known, but if the method approximates a variety of surface
behavior faithfully, then we expect it to give reasonable results in other
cases. Consequently, the implemented methods were tested for accuracy on
artificial data sets constructed by adding Gaussian noise and/or outliers to
different test functions.

\section{Numerical differentiation of noisy data}

Many scientific applications require methods for numerical approximation of
the derivatives of a smooth real-valued function for which only discrete
noise-contaminated scattered data values are available. By scattered data we
understand data which consist of a set of points and corresponding values,
where the points have no structure or order between their positions. Given a
set of $n$ discrete observations $\left( x_{i},y_{i},f_{i}\right) $, the
measurements $f_{i}$ are generally not exact and we assume that they can be
decomposed into two parts $f_{i}=g_{i}+\epsilon _{i}=g\left(
x_{i},y_{i}\right) +\epsilon _{i}$, where $g_{i}$ is \textquotedblleft the
true\textquotedblright\ but unknown deterministic value of the observation
and the error $\epsilon _{i}$ is the statistical uncertainty or departure
from the true inevitably present in the observations. We suppose that the
only source of random variation comes from the unknown variation in the
response, the observed values $f_{i}$ being realizations of some random
variables (i.e., what it is observed are noisy realizations of the true $g$%
). Even if $\left( x_{i},y_{i}\right) $ can be measured with some errors,
they are considered fixed throughout this paper. The points $\left(
x_{i},y_{i}\right) $ are referred to as nodes. For brevity we will use the
notation $\mathbf{x}=\left( x,y\right) $.

If we use conventional finite-difference approximations to estimate the
derivatives then arbitrarily small perturbations in the data can lead to
arbitrarily large variations of the results.

Alternatively, the problem can be addressed as first obtaining a smooth
approximation which may then be differentiated to obtain approximations $%
\hat{g}$ to the desired derivatives. Thus the basic problem is to develop a
statistical model to construct a smooth function $\hat{g}\left( \mathbf{x}%
\right) $ from irregularly spaced data points, to isolate the reliable part
from measurement errors. The estimated function $\hat{g}$ can also be used
to derive new predictions to fill the gaps between the available data. The
properties of the statistical model and of the quantities derived from it
must be studied in an average sense through expectations, variances, and
covariances.

Often the independent variables $\mathbf{x}_{i}=\left( x_{i},y_{i}\right) $
are called explanatory variables, input, predictor, regressor, etc., while
the dependent variable $f_{i}$ the output, outcome, response, etc. The
estimated function $\hat{g}$ is called regression function, smoothing
function, etc. The hat notation\ \textquotedblleft $\hat{\phantom{f}}$%
\textquotedblright\ marks the fit optimal values. This is a standard
notation in statistics, using the hat symbol over a variable to note that it
is a predicted value.

The errors $\epsilon _{i}$ are assumed to be independent and identically
distributed with zero mean $\mathrm{E}\left( \epsilon _{i}\right)=0 $ and
finite variance $\func{var}\left( \epsilon _{i}\right) =\mathrm{E}\left(
\epsilon _{i}^{2}\right) =\sigma ^{2}<\infty $, where $\mathrm{E}\left(
\cdot \right) $ denotes the expected value. According to the central limit
theorem it is reasonable to assume that usually $\epsilon _{i}$ are normal
(Gaussian) random variables. These are statistical errors whose average
tends to zero if enough data are available. In addition, measurements are
also susceptible to systematic errors (like instrument drift during
measurement) that will not diminish with any amount of averaging.

If the measurements were repeated many times then the function $g$ is the
expected function of the statistical model at a fixed value of $\mathbf{x}$: 
$g\left( \mathbf{x}\right) =\mathrm{E}\left[ f\left( \mathbf{x}\right) %
\right] $. As the measurements\ $f_{i}$\ arose from a statistical model and
the errors $\epsilon _{i}$ are random variables, the expected function $\hat{%
g}\left( \mathbf{x}\right) $ can be addressed as a random variable too, and
the mean squared error is: $\mathrm{E}\left[ \left( g\left( \mathbf{x}%
\right) -\hat{g}\left( \mathbf{x}\right) \right) ^{2}\right] =\left( g\left( 
\mathbf{x}\right) -\mathrm{E}\left[ \hat{g}\left( \mathbf{x}\right) \right]
\right) ^{2}+\mathrm{E}\left[ \left( \hat{g}\left( \mathbf{x}\right) -%
\mathrm{E}\left[ \hat{g}\left( \mathbf{x}\right) \right] \right) ^{2}\right]
=\func{bias}\left( \hat{g}\left( \mathbf{x}\right) \right) ^{2}+\func{var}%
\left( \hat{g}\left( \mathbf{x}\right) \right) $. The bias term represents
the amount by which the average of the estimated function $\hat{g}$ differs
from the true value, while the variance term denotes how much $\hat{g}$ will
move around its mean, how much the predictions for a given point vary
between different realizations of the random variable. In order to reduce
the error both terms should be as small as possible, but usually a
compromise is made, the so-called \textquotedblleft bias-variance
tradeoff\textquotedblright . Models that exhibit small variance and high
bias underfit the truth function. Models that exhibit high variance and low
bias overfit the truth function.

\section{Smoothing by local regression}

A simple smoothing method consists of globally fitting the data points with
a polynomial that depends on adjustable parameters. The parameters are
adjusted to achieve a minimum of the function that measures the agreement
between the data and the model. Generally, regression is the process of
fitting models to data using some goodness-of-fit criterion. Polynomials
play an important role in numerical approximation methods because they are
easy to compute, while having the flexibility to approach different
nonlinear relationships. In the case of global methods each approximated
value is influenced by all of the data and hence the use of global methods
is not feasible for a large number of observations because they often
involve the solution of a system $O\left( n\right) $ of equations. As the
number of data increases an increasingly flexible function is needed and
this can be achieved by increasing the degree of the polynomial, but only at
the risk of introducing severe oscillations into the approximate function.

An alternative to global fitting is to keep the polynomial degree low and
build an approximant function by connecting pieces of polynomial functions,
the approximation at any point being achieved by considering only a local
subset of the data. Thus the method does not require the specification of a
global function that characterizes all data, substantially increasing the
domain of surfaces that can be estimated without distortion. The local
methods assume, according to Taylor's theorem, that any function can be well
approximated in a small neighborhood with a low-order polynomial. More
generally, the approximant function satisfying certain conditions can be
chosen from a convenient class of \textquotedblleft
simple\textquotedblright\ parametric functions. In this paper we consider
only linear fitting models, where \textquotedblleft
linear\textquotedblright\ refers to the model's dependence on its fit
parameters, while the functions can be nonlinear on $\left( x,y\right) $. In
fact, local methods involve the use of global methods on smaller sets which
are then \textquotedblleft blended\textquotedblright\ together. In this way $%
\hat{g}$ is obtained by solving many small linear systems of equations
instead of a single but large system, the local methods being capable of
efficiently handling much larger data sets.

Local regression estimation was independently introduced in several
different scientific fields in the late nineteenth and early twentieth
century. In the statistical literature the method was independently
introduced from different viewpoints in the late 1970's. The local
regression belongs to the so called non-parametric regression methods (in
the sense that no parametric form is imposed on the global function), while
the global fitting belongs to the parametric regression methods. The moving
least squares as an approximation method was introduced by Shepard\cite%
{Shepard 1968} in 1968. Another well-known non-parametric regression method
is that of smoothing splines that optimize a penalized least squares
criterion, and so the smoothing splines are also known as penalized least
squares methods. However, smoothing splines require the solution of a global
linear system as they optimize a global criterion and are not generally
local.

The neighborhood of a given point is defined based on some metric. Usually
the Euclidean distance is used, leading to circular neighborhoods, but a
variety of metrics can be achieved by transformation of the original
variables, specifying more general neighborhoods. For example, if the
independent variables are measured on different scales then each variable
can be multiplied by a scale factor and then use the Euclidean distance.
Such a metric would consider the points lying on an ellipse centered at the
given point to be equidistant from the given point.

The size of a neighborhood is given by the so called bandwidth or span, and
it is typically expressed in terms of an adjustable smoothing parameter. We
note that for different methods and/or scientific papers the smoothing
parameter can be defined differently, but ultimately it determines the
neighborhood size. The bandwidth can be fixed or variable as a function of
another parameter. If the nodes are irregularly spaced, for a fixed
bandwidth one may have local estimates based on many points and others only
on few points. Consequently, it is preferable to consider a nearest neighbor
strategy to define the bandwidth, choosing the local neighborhood so that it
always contains a specified number of neighbors.

If the bandwidth is too small then insufficient data are used for estimation
and a noisy fit, or large variance, will result. Conversely, if the
bandwidth is too large then too much data is used and the local estimated
function may not fit the data well and important features of the true
function may be distorted or lost, i.e., the fit will have large bias. The
bandwidth must be chosen to compromise this bias-variance tradeoff. In many
applications is useful to try several different bandwidths as the small
bandwidths may preserve local features that are obscured by larger
bandwidths which, however, may be globally more helpful.

The degree of the local polynomial also affects the bias-variance tradeoff.
A high polynomial degree can provide a better approximation than a low
polynomial degree, leading to an estimate $\hat{g}$ with less bias.

Often in the local fitting a weighting function, also known as kernel
function, is used to give greater weight to points that are closer to the
point whose response is estimated and smaller weight to the points that are
further. Thus the locally weighted regression method combines the kernel
smoothing method with the least square method. The use of the weights is
based on the idea that points near each other are more likely to be related
in a simple way than points that are further apart. The maximum of the
kernel function should be at zero distance, and the function should decay as
the distance increases. Weights that tend to infinity at zero distance allow
exact interpolation, while finite weights lead to smoothing.

A simple weight function raise the distance to a negative power\cite{Shepard
1968}, magnitude of the power determining the rate of decrease of the
weight. These weights tend to infinity at zero, leading to an exact
interpolation. Instead, if the data are noisy a weighting with finite
magnitude is desired. Several types of kernel functions that are commonly
used in the scientific literature and that are implemented in doFORC are
presented in Table Is.\cite{supplementary material} Even if sometimes there
is no clear evidence that the choice of kernel function is critical, there
are examples where one can show differences.\cite{Loader 1999}

For a fitting point $\mathbf{x}$ the bandwidth $h\left( \mathbf{x}\right) $
is defined. For the nearest-neighbor bandwidth the number of nearest
neighbors gives the size of the local neighborhood. If $d$\ is the distance
function and if $W\left( u\right) $\ is a weight function then weight for
the point $\mathbf{x}_{i}$ is $w_{i}=W\left( d\left( \mathbf{x-x}_{i}\right)
\left/ h\left( \mathbf{x}\right) \right. \right) $. The parameters of the
local estimated function are computed in the least squares methods by
minimizing the locally weighted sum of squared deviations between the data
and the model: $\sum\limits_{i\in \mathrm{nearest\ neighbors}}w_{i}\left( 
\mathbf{x}\right) \left[ \hat{g}\left( \mathbf{x}_{i}\right) -f_{i}\right]
^{2}$. The minimization makes no assumptions about the validity of the
model, but it simply finds the best fit to the data.

If the neighboring points are not symmetric about the smoothed point, then
the weight function is not symmetric. This happens near the boundary of the
data set: the kernel for a point near the border is different from the
kernel of an interior point. This effect can lead to poor behavior near
boundaries. This issue should be considered in the analysis of the results.

As the local regression solves a least squares problem, it is a linear
smoother, namely every estimate $\hat{g}_{i}=\hat{g}\left( \mathbf{x}%
_{i}\right) $ is a linear combination of the observed data, and therefore
the vector $\hat{\mathbf{f}}\equiv $ $\hat{\mathbf{g}}=\left( \hat{g}%
_{1},\ldots ,\hat{g}_{n}\right) ^{T}$ of predicted/fitted values at the
observed predictor values is a linear function of the observed/measured
dependent data vector $\mathbf{f}=\left( f_{1},\ldots ,f_{n}\right) ^{T}$: $%
\hat{\mathbf{f}}=L\mathbf{f}$.

The $n\times n$ matrix $L$ is the locally weighted regression matrix, also
known as smoother, influence or hat matrix, and it maps the data to the
predicted values. The $L$ matrix is the counterpart of the orthogonal
projection matrix from the parametric least-squares, but different from it
the $L$ matrix is neither symmetric nor idempotent. The $L$ matrix describes
the influence each response value has on each fitted value. The diagonal
elements of the hat matrix, denoted by $\func{infl}\left( \mathbf{x}%
_{i}\right) $, are called leverages or influences and measure the influence
each response value has on the fitted value for that same observation, or
alternatively the sensitivity of the fitted value to the observed data. The
sum of the diagonal elements, $\func{tr}\left( L \right)$, gives the sum of
the sensitivities with respect to the observed values, and as will be
discussed later, it is one of the degrees of freedom of the model. The hat
matrix can depend on smoothing parameter and $\mathbf{x}$ in a highly
non-linear way, the only linearity in the above equation being the linearity
in $\mathbf{f}$.

\section{Robust smoothing}

An objection to least squares method is lack of robustness to heavy tailed
residual distributions or to the presence of unusual data points in the data
used to fit a model, i.e., to the outliers. An outlier is an observation
that is distant from other observations, having a large residual.
Graphically the outliers are far from the pattern described by the other
points. An outlier may indicate an experimental error, a data entry error or
other problem. We note that the least-squares fitting is a maximum
likelihood estimation of the fitted parameters if the measurement errors are
independent and normally distributed (the principle of maximum likelihood
assumes that the most reasonable values for the smoothing parameters are
those for which the probability of the observed sample is largest). Other
error distributions lead to other fitting criteria. However, even if the
errors are not normally distributed, and then the least-squares estimations
are not maximum likelihood, they may still be useful in a practical sense
for estimating parameters. In many cases the uncertainties associated with a
set of measurements are not known in advance, and then one can assume that
all measurements have the same standard deviation, next fitting for the
model parameters and finally estimating the standard deviation, this
approach at least allowing some kind of error bar to be assigned to the
points.

A simple remedy for outliers is to remove these influential observations
from the least-squares fit.

Another option is to use a fitting method that is more resistant to
outliers, such as one based on the least absolute deviations. However, even
if this method is used the gross outliers can still have a considerable
impact on the model. In addition, working with squares is mathematically
easier than working with absolute values (e.g., it is easier to compute the
derivatives). The least absolute deviations fitting arises as the maximum
likelihood estimate if the errors have a Laplace distribution.

Robust regression methods attempt to remedy the problem by identifying the
influential observations that are suspected of being unreliable and
down-weighting them. There are several ways in which the robustness can be
achieved. The algorithm proposed by Cleveland as part of LOWESS procedure 
\cite{Cleveland 1979} consists in: 1. assign to all observations a
robustness weight $\upsilon _{i}=1$; 2. smooth the data using the weights $%
\upsilon _{i}w_{i}\left( \mathbf{x}\right) $; 3. compute $\hat{\epsilon}%
_{i}=f_{i}-\hat{f}_{i}$ and let $s$ be median of $\left\vert \hat{\epsilon}%
_{i}\right\vert $ (the median is a more robust estimator of the central
value than the mean); 4. assign to the observations the robustness weights $%
\upsilon _{i}=B\left( \left. \hat{\epsilon}_{i}\right/ 6s\right) $, where $B$
is a robustness weight function; 5. repeat steps 2, 3, and 4 until
convergence or by a given number of times. The $B$ function proposed by
Cleveland is the bisquare function.

In all of the following sections, except for that relating to residuals, one
assumes normal (Gaussian) measurement errors. It is important to keep these
limitations in mind, even as we use the very useful methods that follow from
assuming it. Even if the errors are not normally distributed, and therefore
the estimations are not maximum likelihood, they may still be useful in a
practical sense.

\section{Diagnostics and Goodness of fit}

Each smoothing method has one or more smoothing parameters that (together
with the local approximant function and the kernel function) control the
amount of smoothing being performed. When smoothing is accomplished a
question to be asked is how should the smoothing parameters be chosen, how
can the performance of a smoother be evaluated, how well does $\hat{g}$\
estimate the true $g$? Choosing the smoothing parameters is a tradeoff
between small bias and small variance, but more precise characterizations
are needed to derive and study selection procedures. Another more complex
and profound issue is in comparing fits from different smoothers.

Helpful in selecting the smoothing parameter is the visual analysis of the
smoothed data for several smoothing parameters. Nevertheless, this is a
subjective method and consequently objective methods may be preferred in
order to produce an automatic smoothing selection or for consistency of
results among different investigators. Sometimes the selection of fitting
parameters only based on a visual control is critically denoted as
\textquotedblleft chi-by-eye fitting.\textquotedblright

Ideal would be fully automated methods that would be able to provide the
best fit for a given data set. However, this goal is unachievable because
the best fit depends not only on the data set, but also on the questions of
interest.

Nonetheless, what statistics can provide are tools to help guide the choice
of smoothing parameters, to help decide which features are real and which
are random. Even so, no diagnostic technique can provide an unequivocal
answer. Instead, using a combination of diagnostic tools together with the
visual analysis of both the fitted and original data provide insight into
the data. Even if a certain method for selecting the smoothing parameter can
provide a satisfactory result, it is advisable to examine how the fit varies
with the smoothing parameter because sometimes fits with different smoothing
parameters can reveal features that can not be distinguished from just one
\textquotedblleft best\textquotedblright\ fit. There is also the possibility
that fits with very different smoothing parameters can produce similar
goodness of fit, and the use of graphical diagnostics to help make decisions
can become crucial in these cases.

Modeling $f$ comes to a tradeoff between variance and bias. While in some
applications a small variance is preferred, in other applications there is a
strong inclination toward small bias. The advantage of model selection by a
criterion is that it is automated, but it can give a poor solution in a
particular application. For this reason this method has to be combined with
the advantage of graphical diagnostics which allow seeing where the bias is
occurring and where the variability is greatest, allowing us to decide on
the relative importance of each. For example, underestimating a peak in the
surface can be quite undesirable and so we are less likely to accept lower
variance if the result is peak distortion.

\subsection{Residuals}

Residuals are defined as the difference between the observed value and the
predicted value based on the regression, measuring the discrepancy between
the data and the fitted model: $\hat{\epsilon}_{i}=f_{i}-\hat{f}_{i}$. While
the error $\epsilon _{i}$ is the deviation of the observed value from the
unknown true value, the residual $\hat{\epsilon}_{i}$ is the difference
between the observed value and the estimated value. Residuals can be used to
check assumptions on the statistical errors. If $I$\ is the $n\times n$
identity matrix then the vector of residuals is $\hat{\mathbf{\epsilon }}%
=(I-L)\mathrm{f}$. Residual sum of squares (RSS), also known as sum of
squared residuals (SSR) or sum of squared errors of prediction (SSE), is the
sum of the squares of residuals: $\mathrm{RSS}=\hat{\mathbf{\epsilon }}^{T}\,%
\hat{\mathbf{\epsilon }}=\sum\nolimits_{i=1}^{n}\hat{\epsilon}_{i}^{2}$. It
is a global measure of the discrepancy between the data and an estimation
model. A small RSS indicates a tight fit of the model to the data and
possibly overfitting (undersmoothing) of data, which essentially means that
the predicted function $\hat{g}$ is sensitive to the noise, following the
data too closely. In the least squares method the smoothing coefficients are
chosen so as to minimize RSS.

Graphical tools to help in selecting the smoothing parameter are: (i)
residuals vs. independent (predictor) variables to detect lack of fit; if
the residuals exhibit a pattern or structure then the corresponding
smoothing parameter value may not be satisfactory; (ii) residuals vs. fitted
values to detect a dependence of the scale of the errors on the level of the
fitted values; (iii) sequential plot of residuals in the order the data were
obtained to detect a possible shift over time.

Generally, as the smoothing parameter is reduced, the residuals become
smaller and exhibit less structure.

It is also helpful to smooth the residual plots to enhance the observer view
of residuals, to assist in discerning patterns or clusters in the residual
plots. This also facilitates to determine if the residuals are symmetrical
around zero, to detect bias. By reducing the noise of the residual plots,
the attention can be attracted more easily to features that have been missed
or not properly modeled by the smooth. The residual plots should be used in
conjunction with the plots of the observed and predicted data to determine
if large residuals correspond to features that have been inadequately
modeled.

doFORC contains these graphical tools necessary to analyze the numerical
results.

However, the residual analysis does not provide an indication of how well
the model will make new predictions to fill the gaps between the available
data. In the case of overfitting the function performs well on the given set
of data, but because it have a tendency to be too close to this data set it
may have poor predictive precision. One way to overcome this issue is to use
cross-validation to measure the predictive performance of the model.

\subsection{Cross-validation}

Cross-validation (CV) focuses on the prediction problem: using the
regression function to predict new observations, how good will the
prediction be? The intuitive idea of cross-validation is to split the
observations in two sets. The model is fitted using only the first set and
then it is used to obtain predictions for the observations from the second
set. A single data split provides a validation estimate, and averaging over
several splits provides a cross-validation estimate. Various splitting
strategies lead to various cross-validation estimates. One of the most
common is leave-one-out cross validation (LOOCV) for which each data point
is successively omitted and used for validation. The idea of LOOCV is that
the best model for the measurements is the model that best predicts each
measurement as a function of the others. If the number of observations is
large the LOOCV estimate seems to be computationally expensive requiring $n$
model fits. However, the linear regression models require only one fit over
the entire data set of observations. The Generalized Cross-Validation
criterion (GCV), proposed in the context of smoothing splines by Craven and
Wahba,\cite{GCV: Craven 1979} is an approximation to the LOOCV and it
requires only finding the trace of the hat matrix: $\mathrm{GCV}=n\hat{\sigma%
}^{2}\left/ \left( n-\func{tr}\left( L\right) \right) ^{2}\right. $, where $%
\hat{\sigma}^{2}\equiv \mathrm{RSSm}=\mathrm{RSS}/n$. The model that
minimizes the value of GCV over a suitable range can be selected as the
optimal model for the data.

\subsection{Information criteria}

Taking into account that the true model is unknown, it is attempted to
quantify the information that is lost by using an approximate model on the
available data. There are several information criteria, derived from
different theoretical considerations. Two common criteria are the AIC
(Akaike Information Criterion)\cite{AIC: Akaike 1973, AIC: Akaike 1974, AIC:
Akaike} and BIC (Bayesian Information Criterion).\cite{Schwarz 1978}

Given a collection of models (of values of the smoothing factor in our case)
for the data, AIC estimates the quality of each model, relative to each of
the other models. If the data are generated by the function $g$ and we
consider two candidate models $\hat{g}_{1}$ and $\hat{g}_{2}$ to represent $%
g $, then the information lost from using $\hat{g}_{1}$ to represent $g$ can
be found by calculating the corresponding Kullback--Leibler discrepancy
function,\cite{Kullback-Leibler 1951} and similar for $\hat{g}_{2}$ . If we
know $g$ then we would choose the model that minimizes the information loss.
As we do not actually know $g$, Akaike shown that instead we can estimate
via AIC how much more (or less) information is lost by $\hat{g}_{1}$ than by 
$\hat{g}_{2}$.

AIC does not provide information about the absolute quality of a single
model, only the quality relative to other models, being useful for comparing
models. If all the candidate models fit poorly then AIC does not provide a
warning. In itself, the value of AIC has no meaning. For large $n$
minimizing AIC is asymptotically equivalent to minimizing GCV.

However, AIC might lead to variable choices of smoothing parameter and to
overfitting if the number of observations is small. Hurvich et al.\cite%
{AICC: Hurvich 1998} added a correction factor and developed two criteria $%
\mathrm{AIC}_{\mathrm{C}_{0}}$ and $\mathrm{AIC}_{\mathrm{C}_{1}}$. The
foremost is the more exact of the two, but requires numerical integration
for its evaluation. $\mathrm{AIC}_{\mathrm{C}_{1}}$ is an approximation to $%
\mathrm{AIC}_{\mathrm{C}_{0}}$. Because $\mathrm{AIC}_{\mathrm{C}_{1}}$
still requires calculations involving all the elements of a $n\times n$
matrix, was developed $\mathrm{AIC}_{\mathrm{C}}$, which is an approximation
to $\mathrm{AIC}_{\mathrm{C}_{1}}$ that is as simple to apply as AIC as it
is a function of $\func{tr}\left( L\right) $ : $\mathrm{AIC}_{\mathrm{C}%
}=\log \left( \hat{\sigma}^{2}\right) +1+2\left( \func{tr}\left( L\right)
+1\right) \left/ \left( n-\func{tr}\left( L\right) -2\right) \right. $ and $%
\mathrm{AIC}_{\mathrm{C}_{1}}=\log \left( \hat{\sigma}^{2}\right) +n\left(
\delta _{1}\left/ \delta _{2}\right. \right) \left. \left( n+\func{tr}\left(
L^{T}\,L\right) \right) \right/ \left( \delta _{1}^{2}\left/ \delta
_{2}\right. -2\right) $, where $\delta _{1}=\func{tr}\left( I-L\right)
^{T}\,\left( I-L\right) $ and $\delta _{2}=\func{tr}\left( \left( I-L\right)
^{T}\,\left( I-L\right) \right) ^{2}$.

\subsection{Degrees of freedom}

In order to compare different amounts of smoothing being performed by
different smoothing methods (e.g., quadratic versus cubic polynomials) the
smoothers should be placed on an equal footing. Using the same bandwidth
does not give rise to meaningful results as, for example, a higher order fit
is more variable but less biased. Instead, number of degrees of freedom of a
smoother, also called effective number of parameters, is an indication of
the amount of smoothing. To compare different smoothers we simply choose
smoothing parameters producing the same number of degrees of freedom.
Degrees of freedom\cite{DF: Hastie 1990} (DF)in a nonparametric fit is a
number that is analogous to the number of fit parameters in a parametric
model. For a linear nonparametric smoother there are three commonly used
measures of model degrees of freedom to provide a quantitative measure of
the estimator complexity: $\mathrm{DF1}=\func{tr}\left( L\right) $, $\mathrm{%
DF2}=\func{tr}\left( L^{T}\,L\right) $, $\mathrm{DF3}=2\func{tr}\left(
L\right) -\func{tr}\left( L^{T}\,L\right) =2\mathrm{DF1}-\mathrm{DF2}%
=n-\delta _{1}$, and they are not necessarily integer numbers These
definitions can be motivated by analogy with linear parametric regression
models and are useful for different purposes. DF1 is the easiest to compute.
DF2 is as well referred to as the equivalent number of parameters.

In the case of the parametric least-squares the trace of the orthogonal
projection matrix (which is the counterpart of the hat $L$ matrix) gives the
dimension of the projection space, which is also the number of basis
functions, and hence the number of parameters involved in the fit.

More smoothing means a more restricted function $\hat{g}$, namely fewer
degrees of freedom, i.e., the degrees of freedom is a quantity that
summarizes the flexibility of $\hat{g}$. The usefulness of the degrees of
freedom is in providing a measure of the amount of smoothing that is
comparable between different estimates applied to the same dataset.

As for a parametric regression $L$ is symmetric and idempotent the above
definitions coincide and usually equal the number of parameters. For
nonparametric regression the definitions are usually not equal and $1\leq 
\mathrm{DF2}\leq \mathrm{DF1}\leq \mathrm{DF3}\leq n$.

\subsection{Statistical properties}

As long as $\hat{g}$ is only an estimate of the true function $g$, it is
essential to examine the results statistically.

As we have assumed that all noise terms $\epsilon _{i}$ have the normal
distribution $N\left( 0,\sigma ^{2}\right) $ and are independent of each
other and of $\mathbf{x}$, the vector $\mathbf{\epsilon }$ of all noise
terms has a multivariate normal distribution with mean vector $\mathbf{0}$,
and variance matrix $\sigma ^{2}I$. As the estimated $\hat{\mathbf{f}}$ is
function of $\mathbf{f}$, the statistical fluctuations in $\mathbf{f}$ leads
to statistical fluctuations in $\hat{\mathbf{f}}$. Accordingly the vector of
estimated values and the vector of residuals are each multivariate normal
distributions, inheriting their distributions from that of the data, namely
from the noise term.

The quantity of interest is the vector $\mathbf{\mu }=\left( \mu _{1},\ldots
,\mu _{n}\right) ^{T}$, with $\mu _{i}=g\left( \mathbf{x}_{i}\right) $, of
the unknown regression function evaluated at points $\mathbf{x}_{i}$, and it
is the conditional expectation of $\mathbf{f}$: $\mathrm{E}\left( \mathbf{f}%
\right) =\mathbf{\mu }$. The estimate of $\mathbf{\mu }$ is the fitted
vector $\hat{\mathbf{f}}$, and the expectation of this estimate is: $\mathrm{%
E}(\hat{\mathbf{f}})=\mathrm{E}\left( L\mathbf{f}\right) =L\mathrm{E}\left( 
\mathbf{f}\right) =L\mathbf{\mu }$. The difference is the bias of the
estimator $\mathbf{b=}\mathrm{E}\left( \mathbf{f}\right) -\mathrm{E}(\hat{%
\mathbf{f}})=\left( I-L\right) \mathbf{\mu }$, and it depends on the unknown 
$\mathbf{\mu }$. Generally smoothers are biased, i.e., $\mathbf{b\neq 0}$.

As the estimated value $\hat{f}_{i}$ is a linear combination of the
individual observations $f_{i}$, $\hat{f}_{i}=\sum\nolimits_{j=1}^{n}l_{ij}%
\,f_{j}$, where $\mathbf{l}_{i}=\left( l_{i\,1},\ldots ,l_{i\,n}\right) $ is
the $i^{\mathrm{th}}$ row of the hat matrix $L$, and as the observed value $%
f_{i}$ varies about the true value with variance $\sigma ^{2}$, the
estimated $\hat{f}_{i}$ has variance $\func{var}(\hat{f}_{i})=\sum%
\nolimits_{j=1}^{n}l_{ij}^{2}\,\,\func{var}\left( f_{j}\right) =\sigma
^{2}\left\Vert \mathbf{l}_{i}\right\Vert ^{2}$. Note the distinction between
the variance $\sigma ^{2}$ of the observation $f_{i}$ and the variance of
the estimated $\hat{f}_{i}$. Due to the measurement errors the observed
vector $\mathbf{f}=\left( f_{1},\ldots ,f_{n}\right) ^{T}$ is only one of
the possible realizations of the true value $\mathbf{\mu }$, and there are
infinitely many other realizations, each of which could have been the one
measured, but happened not to be. If we take additional sets of $n$
measurements each, performed under the same conditions, we would obtain a
set of estimated values $\hat{\mathbf{f}}$ that will be clustered about $%
\mathrm{E}(\hat{\mathbf{f}})$, but with a distribution that is narrower than
the distribution of $\mathbf{f}$. The quantity $\left\Vert \mathbf{l}%
_{i}\right\Vert ^{2}$ measures the variance reduction of the smoother at
data point $\mathbf{x}_{i}$ due to the local regression. A global measure of
the amount of smoothing is provided by $\sum\nolimits_{j=1}^{n}\left\Vert 
\mathbf{l}_{i}\right\Vert ^{2}=\func{tr}\left( L^{T}\,L\right) =\mathrm{DF2}$%
, called also degrees of freedom for variance. As the amount of smoothing
increases, $\func{tr}\left( L^{T}\,L\right) $ tends to decrease, while the
elements of bias $\mathbf{b}$ tend to increase, and conversely. The variance
of the data vector $\mathbf{f}$ is the variance of the noise vector $\mathbf{%
\epsilon }$, i.e., $\func{var}\left( \mathbf{f}\right) =\sigma ^{2}I$ and
accordingly the variance of the estimate is $\func{var}(\hat{\mathbf{f}}%
)=\sigma ^{2}LL^{T}$. Roughly, two fits with the same degrees of freedom
have the same variance.

The variance of the residual vector is $\func{var}(\mathbf{f}-\hat{\mathbf{f}%
})=\sigma ^{2}\left( I-L\right) \left( I-L\right) ^{T}=\sigma ^{2}\left(
I-L-L^{T}+LL^{T}\right) $. In order to obtain an unbiased estimate of the
unknown $\sigma ^{2}$, in the case of the parametric least squares method
the residual sum of squares $\mathrm{RSS}=\left\Vert \mathbf{f}-\hat{\mathbf{%
f}}\right\Vert ^{2}=\sum\nolimits_{i=1}^{n}(f_{i}-\hat{f}_{i})^{2}$ is
divided by the degrees of freedom for error. Instead, in general smoothers
are biased and RSS has expectation $\mathrm{E}\left( \mathrm{RSS}\right)
=\sigma ^{2}\func{tr}\left( I-L-L^{T}+LL^{T}\right) +\mathbf{b^{T}b}=\sigma
^{2}\left( n-\mathrm{DF3}\right) +\mathbf{b^{T}b}$. A biased estimate of $%
\sigma ^{2}$ can be obtained ignoring the bias term in the above equation: $%
\hat{\sigma}^{2}=\mathrm{RSS}/\left( n-\mathrm{DF3}\right) =\mathrm{RSS}%
/\delta _{1}=\mathrm{RSE}^{2}$, which overestimates $\sigma ^{2}$. The
quantity $\delta _{1}=n-\mathrm{DF3}$ is called the degrees of freedom for
error of the smoother, since in the linear regression this is $n-p$, where $%
p $ is the number of parameters. The estimate $\hat{\sigma}$ is also called
the residual standard error (RSE).

In choosing the smoothing parameter, we need not try to minimize the mean
squared error at each $\mathbf{x}_{i}$, but instead we have to focus on a
global measure as the average mean squared error: $\mathrm{MSE}=\left(
1/n\right) \sum\nolimits_{i=1}^{n}\mathrm{E}\left( (f_{i}-\hat{f}%
_{i})^{2}\right) =\left( 1/n\right) \mathbf{b^{T}b+}\left( 1/n\right) \sigma
^{2}\func{tr}\left( L^{T}\,L\right) =\func{bias}+\func{variance}$. It would
be desirable that both terms be as small as possible. If the bias is big
then the estimate $\hat{\mathbf{f}}$ is off center, while if the variance is
big then the estimate $\hat{\mathbf{f}}$ is too variable (its distribution
has too much spread). Reducing smoothing reduces bias but increases
variance, and vice versa (bias-variance tradeoff). Generalized cross
validation GCV can be mathematically justified in that asymptotically it
minimizes mean squared error for the estimation of $\mathbf{\mu }$.

\subsection{Confidence intervals}

Given an estimate of $\sigma ^{2}$, the diagonal of the variance matrix of
the fitted vector $\hat{\mathbf{f}}$ can be used to form point-wise standard
error bands for the true mean $\mathbf{\mu }$. For a negligible bias these
bands represent point-wise confidence intervals. If the bias is not
negligible (which is very difficult to check), then the bands provide a
point-wise confidence interval for $L\mathbf{\mu }$ rather than for $\mathbf{%
\mu }$. Inference for $\sigma ^{2}$ is based on the distribution of $\hat{%
\sigma}^{2}$ (statistical inference being the process of using data analysis
to deduce properties of an underlying probability distribution in the
presence of uncertainty, the process of drawing conclusions about population
parameters based on a sample taken from the population).

A local estimate $\hat{f}_{i}$ has the distribution $(\hat{f}_{i}-\mu
_{i})\left/ \sigma \left\Vert \mathbf{l}_{i}\right\Vert \right. \sim N\left(
0,1\right) $. If the estimate is unbiased, so that $\mathrm{E}(\hat{f}%
_{i})=\mu _{i}$, confidence interval may take the form $\hat{f}_{i}\pm
c\sigma \left\Vert \mathbf{l}_{i}\right\Vert $, where the coefficient $c$ is
chosen as the $1-\alpha /2$ quantile of the standard normal distribution $%
N\left( 0,1\right) $: $P\left( \left\vert \hat{f}_{i}-\mathrm{E}(\hat{f}%
_{i})\right\vert <c\sigma \left\Vert \mathbf{l}_{i}\right\Vert \right)
=1-\alpha $, where $0<\alpha <1$ is the desired significance level, and $%
P(\cdot )$ refers to the probability that an event will occur.

When $\sigma $ is replaced by the residual standard deviation $\hat{\sigma}$
then as a first approximation the random variable $(\hat{f}_{i}-\mu
_{i})\left/ \hat{\sigma}\left\Vert \mathbf{l}_{i}\right\Vert \right. $ can
be approximated by a Student's $t\mathrm{-distribution}$ with $\delta _{1}=n-%
\mathrm{DF3}$ degrees of freedom, and accordingly $\left. \delta _{1}\hat{%
\sigma}^{2}\right/ \sigma ^{2}$ is approximated by a $\chi ^{2}$
distribution with $\delta _{1}$ degrees of freedom. This approximation leads
to the use of the quantiles of the $t\mathrm{-distribution}$ with $\delta
_{1}$ degrees of freedom.

A better approximation can be obtained through a two-moment correction, this
approximation leading to the use of the percentiles of the $t\mathrm{%
-distribution}$ with $\rho =\delta _{1}^{2}\left/ \delta _{2}\right. $
degrees of freedom, called also lookup degrees of freedom.

\subsection{Automatic smoothing parameter selection}

There are several methodologies for automatic smoothing parameter selection
that generally fall into two broad classes of methods:

-- first class of methods selects the value of that smoothing parameter that
minimizes a criterion that incorporates both the tightness of the fit and
model complexity. Such a criterion can usually be written as the sum of a
function that estimates the mean average squared error and a penalty
function designed to decrease with increasing smoothness of the fit. The
first term measures the goodness of fit while the second term controls for
model complexity. This trading between goodness of fit and complexity may
also be seen as the trade-off between bias and variance. The GCV, $\mathrm{%
AIC}_{\mathrm{C}}$, and $\mathrm{AIC}_{\mathrm{C}_{1}}$ criteria belong to
this class of methods.

-- the second class of methods attempts to set an approximate measure of
model degrees of freedom to a specified target value. These methods are
useful for making meaningful comparisons between different fits. The DF1,
DF2, and DF3 are three approximate model degrees of freedom for a model.

In terms of computational effort $\mathrm{AIC}_{\mathrm{C}}$, GCV, and DF1
depend on the smoothing matrix $L$ only through its trace. In contrast, $%
\mathrm{AIC}_{\mathrm{C}_{1}}$, DF2, and DF3 depend on the entire $L$ matrix
and accordingly the time taken to compute these quantities may dominate the
time required for the model fitting.

\section{LOESS method}

LOWESS (LOcally Weighted Scatterplot Smoothing) method proposed in 1979 by
Cleveland\cite{Cleveland 1979} was designed for univariate (one independent
variable) weighted regression of scattered points using a nearest neighbors
bandwidth, local linear polynomials, and a tricube kernel. However, any
other weight function that satisfies the properties presented in Ref. %
\onlinecite{Cleveland 1979} could be used. In addition, the method proposes
the use of a robust regression within each neighborhood, with a bisquare
robustness weight function, to protect against outliers.

LOESS (LOcal regrESSion) proposed in 1988 by Cleveland and Devlin\cite%
{Cleveland 1988} generalized the initial method to the multivariate case and
allows the user to choose linear or second-order polynomials for the local
fitting. Statistical properties of the method are analyzed as well. As $%
\delta _{1}$ and $\delta _{2}$ require calculations involving all the
elements of the hat matrix, in Ref. \onlinecite{Cleveland 1991}
approximations based on the trace of the matrix are provided for them.

The method has been implemented in FORTRAN 77 and is available as open
source.\cite{LOESS} In solving the least-squares problem a preliminary QR
decomposition followed by the singular-value decomposition of the triangular
matrix R allows the pseudo-inverse to be computed efficiently. The program
allows users to incorporate extra nonnegative constants, or weights,
associated with each data point, into the fitting criterion. The size of the
weight indicates the precision of the information contained in the
associated observation.

\section{Modified Shepard methods}

Original Shepard method\cite{Shepard 1968} is a global interpolation method
based on a weighted average of values $f_{k}$ at data points, with the
weight functions given by the inverse of the squares of the distance between
the data points. The algorithm was later enhanced to address some of its
shortcomings. The method was modified to become local by Franke and Nielson%
\cite{Franke-Nielson 1980} by using weight functions with local support.
Modified Shepard method also replaces $f_{k}$ in the weighted average with
suitable local approximations $P_{k}(\mathbf{x})$ that interpolates the data
value at node $k$ and locally fits the data values on a set of nearby nodes
in a weighted least-squares sense. Quadratic polynomial functions are used
for $P_{k}$. The interpolant function is defined as a convex combination of
the nodal functions $P_{k}$: $F(\mathbf{x})=\sum\nolimits_{k=1}^{n}W_{k}(%
\mathbf{x})P_{k}(\mathbf{x})\left/ \sum\nolimits_{i=1}^{n}W_{i}(\mathbf{x}%
)\right. $, with the weight functions taken as $W_{k}(\mathbf{x})=\left[
\left. \left( R_{w}-d_{k}\right) _{+}\right/ \left( R_{w}d_{k}\right) \right]
$, where $\left( R_{w}-d_{k}\right) _{+}=R_{w}-d_{k}$ if $d_{k}<R_{w}$, and
equals 0 otherwise, $d_{k}(\mathbf{x})$ is the Euclidean distance between $%
\mathbf{x}_{k}$ and $\mathbf{x}$, $R_{w}$ is a radius of influence about $%
\mathbf{x}_{k}$, and the exponent parameter usually $u=2$ although other
values may be used.

The method was improved by Renka et al.,\cite{Renka 1988 Q, Renka 1999 C,
Renka 1999 T} which also developed several variations of the method based on
polynomial and trigonometric functions for $P_{k}$ in order to increase the
precision of the approximation: quadratic polynomial\cite{Renka 1988 Q} $%
P_{k}(x,y)=a_{1k}\left( x-x_{k}\right) ^{2}+a_{2k}\left( x-x_{k}\right)
\left( y-y_{k}\right) +a_{3k}\left( y-y_{k}\right) ^{2}+a_{4k}\left(
x-x_{k}\right) +a_{5k}\left( y-y_{k}\right) +f_{k}$; cubic polynomial\cite%
{Renka 1999 C} $P_{k}(x,y)=a_{1k}\left( x-x_{k}\right) ^{3}+a_{2k}\left(
x-x_{k}\right) ^{2}\left( y-y_{k}\right) +a_{3k}\left( x-x_{k}\right) \left(
y-y_{k}\right) ^{2}+a_{4k}\left( y-y_{k}\right) ^{3}+a_{5k}\left(
x-x_{k}\right) ^{2}+a_{6k}\left( x-x_{k}\right) \left( y-y_{k}\right)
+a_{7k}\left( y-y_{k}\right) ^{2}+a_{8k}\left( x-x_{k}\right) +a_{9k}\left(
y-y_{k}\right) +f_{k}$; \ cosine series\cite{Renka 1999 T} $%
P_{k}(x,y)=a_{1k}+a_{2k}\cos (p)+a_{3k}\cos (q)+a_{4k}\cos (2p)+a_{5k}\cos
(p)\cos (q)+a_{6k}\cos (2q)+a_{7k}\cos (3p)+a_{8k}\cos (2p)\cos
(q)+a_{9k}\cos (p)\cos (2q)+a_{10k}\cos (3q)$, with $p=\left( x-x_{\min
}\right) \left/ \left( x_{\max }-x_{\min }\right) \pi \right. $, $q=\left(
y-y_{\min }\right) \left/ \left( y_{\max }-y_{\min }\right) \pi \right. $,
where $x_{\min }$, $x_{\max }$, $y_{\min }$, $y_{\max }$ are the extremes of
the nodal coordinates. The power parameter $u=2$ for the quadratic version,
while $u=3$ for the other two methods. The complexity of the nodal functions
determines their potential to track data. Cubic and trigonometric
approximations tend to perform better in situations where the surface has
substantial curvature, such as local sharp maxima and/or minima. The
interpolant function is of class $C^{1}$ for the quadratic method and of
class $C^{2}$ for the cubic and trigonometric methods.

The coefficients of the nodal functions are chosen to minimize the weighted
square error with the kernel function $\left[ \left. \left( R_{\mathrm{%
influence}}-d_{ik}\right) _{+}\right/ \left( R_{\mathrm{influence}%
}d_{ik}\right) \right] $, where $d_{ik}$ is the Euclidean distance between
nodes $i$ and $k$, and $R_{\mathrm{influence}}$ is a radius of influence
about node $k$ within which data is used for the least square fit. $R_{%
\mathrm{influence}}$ is denoted by $R_{q}$ for the quadratic version and by $%
R_{c}$ for the other two methods. While the Franke-Nielson method uses fixed 
$R_{w}$ and $R_{q}$, the Renka methods allow these radii to vary with $k$,
choosing the radii so as to include $N_{w}$ and $N_{q},N_{c}$ nodes
respectively, with fixed values of $N_{w},N_{q},N_{c}$. The $N_{w}$
parameter controls number of nodal functions used to build the interpolant
function. The $N_{q},N_{c}$ parameters control the number of nodes used to
build nodal functions.

The modified Shepard methods were implemented by Renka et al. in FORTRAN 77
and are available as open source.\cite{Shepard} The efficient nearest
neighbors search algorithm leads to good performance even for large
datasets. The solution of the least-squares problem is obtained via the QR
decomposition computed with a series of Givens rotations. This requires only
a $6\times 6$\ workspace array in the case of a quadratic or a $10\times 10$
array otherwise. The triangular matrix $R$ is built up gradually as the rows
of the least squares problem are processed one by one in the order of
nearest neighbors to some node $k$, and as a result the method is simple and
computationally very efficient as it does not require to solve linear
equations, making the solution of very large problems computationally
feasible with respect to execution time and memory requirements.

We have generalized the modified Shepard methods to smoothing methods, which
are appropriate when the data are inaccurate or noisy. For this we have
removed the constraint that the nodal functions interpolate the data values,
i.e., $P_{k}(x,y)\neq f_{k}$. In this case $N_{q},N_{c}$ control also the
noise suppression (smoothness): $N_{q},N_{c}$ noisy points are combined to
get a noise-free surface and consequently the larger level of noise is, the
larger $N_{q},N_{c}$ are needed.

\section{doFORC tool software}

The main features of the doFORC tool\cite{doFORC} are as follows:

-- is a portable (standalone) application working on various operating
systems, is made using only free libraries, and it is made freely available
to the scientific community

-- the user easy to use graphical user interface (GUI) allows users to
import the input data, set the fitting parameters, to graphical represent
(2D, 3D, and projection) both input and output data, and to export the
graphs to image files

-- allows the choice of one of the four implemented nonparametric regression
procedures: LOESS and three modified Shepard methods further modified for
noisy data. The procedures allow great flexibility because no assumptions
about the parametric form of the regression surface are needed. Thus the
users can try different methods using their data and select one (or more)
that is best suited to their needs.

-- allows the use of different kernel functions

-- input data may have various formats, including the PMC MicroMag format
for which the drift correction can be performed

-- allows the use of user weights associated with each data point, weights
that indicates the precision of the information contained in the associated
observation

-- removes points that are closer than some tolerance (duplicate or nearby
points) from the input data

-- input data can be cropped to ignore certain parts of the input points

-- allows the use of a scale factor for input data to change the shape of
the neighborhood, considering the points lying on an ellipse centered at the
given point to be equidistant from the given point. This feature is useful
when the variables have different scales.

-- allows the standardization of the input data to change the shape of the
neighborhood. This feature is useful when the variables have significantly
different scales. The standardization is accomplished using Winsorized mean
and standard deviation of each variable. Winsorized values are robust scale
estimators in that extreme values of a variable are discarded (the smallest
and largest 5\% of the data) before estimating the data scaling.

-- robust smoothing allows to minimize the influence of outlying data
(outliers)

-- output data can be provided at the user's choice: (i) in the input
points, (ii) in a regular grid in the $\left( h_{\mathrm{applied}}\geq h_{%
\mathrm{reversal}},h_{\mathrm{reversal}}\right) $ half-plane, (iii) in a
regular grid in the $\left( h_{\mathrm{coercice}}\geq 0,h_{\mathrm{%
interaction}}\right) $ half-plane, (iv) in a general rectangular regular
grid, (v) or in the user points

-- output consist in: (i) predicted smoothed values in the points from the
input file, along with other information (residuals, smoothed residuals,
histogram of the residuals, influences, confidence intervals for fit, radius
of influence), (ii) requested derivatives in the output points, (iii)
requested statistics

-- performs statistical inference (deduces properties of data sets from a
set of observations and hypotheses) provided that the error distribution
satisfies some basic assumptions

-- in order to perform the diagnostics and goodness of fit doFORC compute
the residuals to characterize the difference between the actual observed
value and the predicted value, generalized cross-validation GCV to measure
the predictive performance of the model, two information criteria $\mathrm{%
AIC}_{\mathrm{C}}$ and $\mathrm{AIC}_{\mathrm{C}_{1}}$ to quantify the
information that is lost by using an approximate model on the available
data, and three degrees of freedom DF1, DF2, DF3 to compare different
amounts of smoothing being performed by different smoothing methods.

-- based on the above criteria doFORC can perform automatic smoothing
parameter selection. Although the default method for selecting the smoothing
parameter value is often satisfactory, it is often a good practice to
examine how the fit varies with the smoothing parameter. In some cases, fits
with different smoothing parameters might reveal important features of the
data that cannot be discerned by looking at a fit with just a single
\textquotedblleft best\textquotedblright\ smoothing parameter.

-- there are several ways in which user can control the sequence of fitting
parameters (number of neighbors ) examined:

\begin{itemize}
\item \setlength{\itemsep}{0em}\setlength{\parskip}{0em}\setlength{%
\parsep}{0em}\setlength{\topsep}{0em}\setlength{\partopsep}{0em}%
\setlength{\parindent}{0em}

\item[(i)] specifying a list of $nn$ values and (i$_{1}$) if no criterion is
specified then a separate fit is provided for each $nn$ value, while (i$_{2}$%
) if a criterion is specified then all values specified in $nn$ list are
examined and the value that minimizes the specified criterion is selected

\item[(ii)] specifying a range $(lower,upper)$ of $nn$ values examined for
which the golden section search method is used to find a local minimum of
the specified criterion in the given range
\end{itemize}

-- provides test problems that consist of sets of data obtained using
various known functions, over which a known normal (Gaussian) noise and a
certain percentage of outliers are added. These test problems embedded in a
dedicated GUI allow users to see the limits of each method, to observe any
numerical artifacts. The test problems can also be used to test, asses the
accuracy, and validate other FORC type (or two dimensional smoothing)
software tools that exist in the scientific literature.

\section{Test functions results}

\begin{figure*}[tbp]
\includegraphics[width=180mm,keepaspectratio=true]{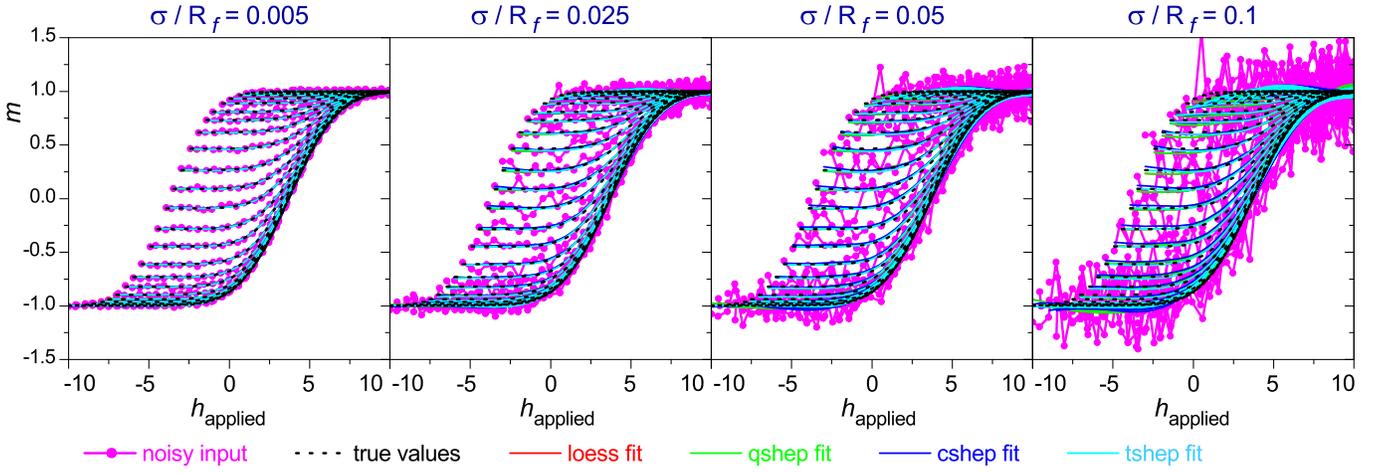}
\caption{The \textquotedblleft true\textquotedblright regression $f_{3}$
function, the noisy input, and the fitted values using automatic smoothing
parameter selection with the $\mathrm{AIC_{C}}$ criterion, for different
values of the standard deviations $\protect\sigma \left/ R_{f}\right. $.}
\label{Fig_1}
\end{figure*}

\begin{figure*}[tbp]
\includegraphics[width=180mm,keepaspectratio=true]{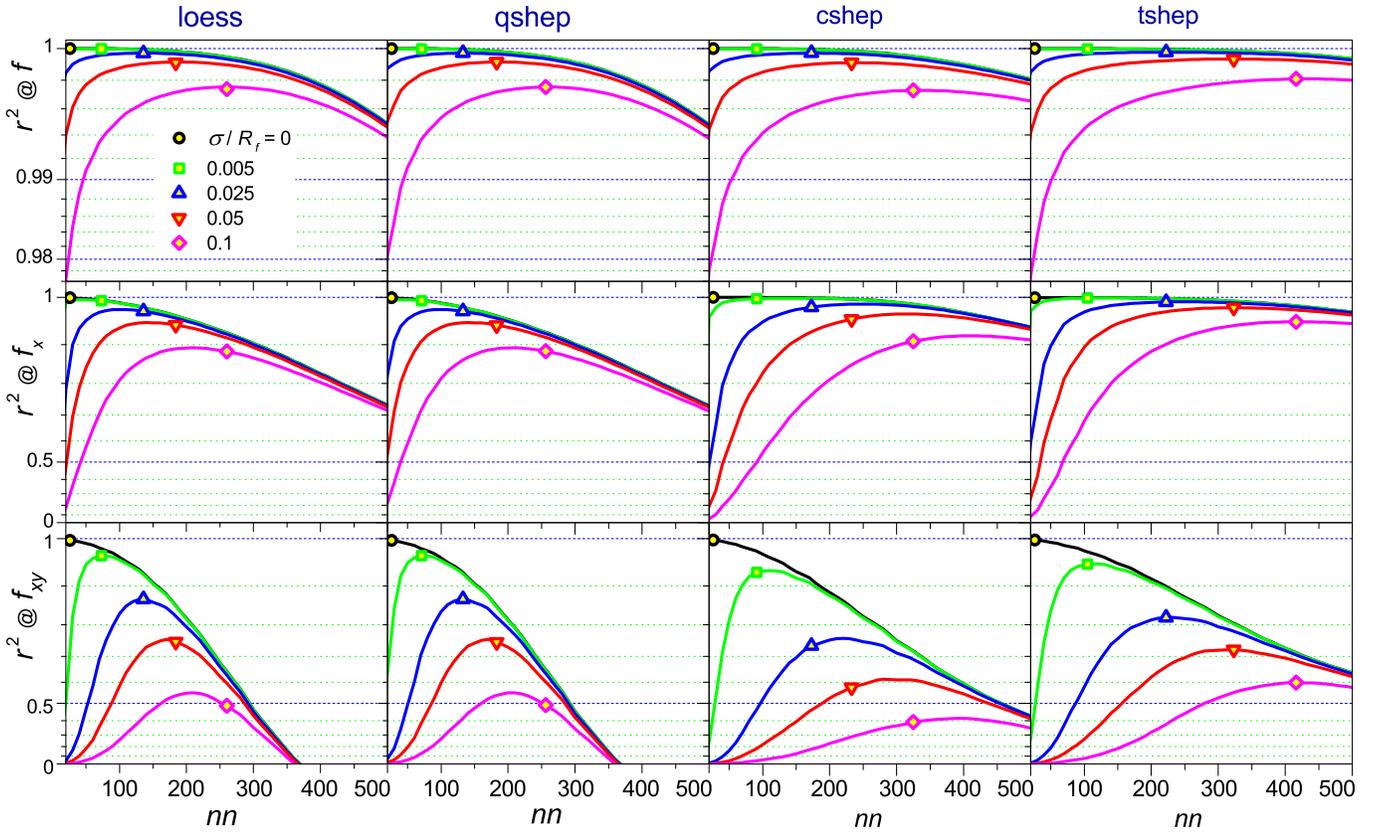}
\caption{Dependence of the coefficient of determination $r^2$ on the on the
number $nn$ of nearest neighbor data points used for the local nonparametric
regression, for test function $f_{3}$ and for its first derivative $f_{x}$
and for the mixed derivative $f_{xy}$, for different amplitudes of the
Gaussian noise. The symbols mark the values given by the automatic smoothing
parameter selection with the $\mathrm{AIC_{C}}$ criterion.}
\label{Fig_2}
\end{figure*}

\begin{figure*}[tbp]
\includegraphics[width=180mm,keepaspectratio=true]{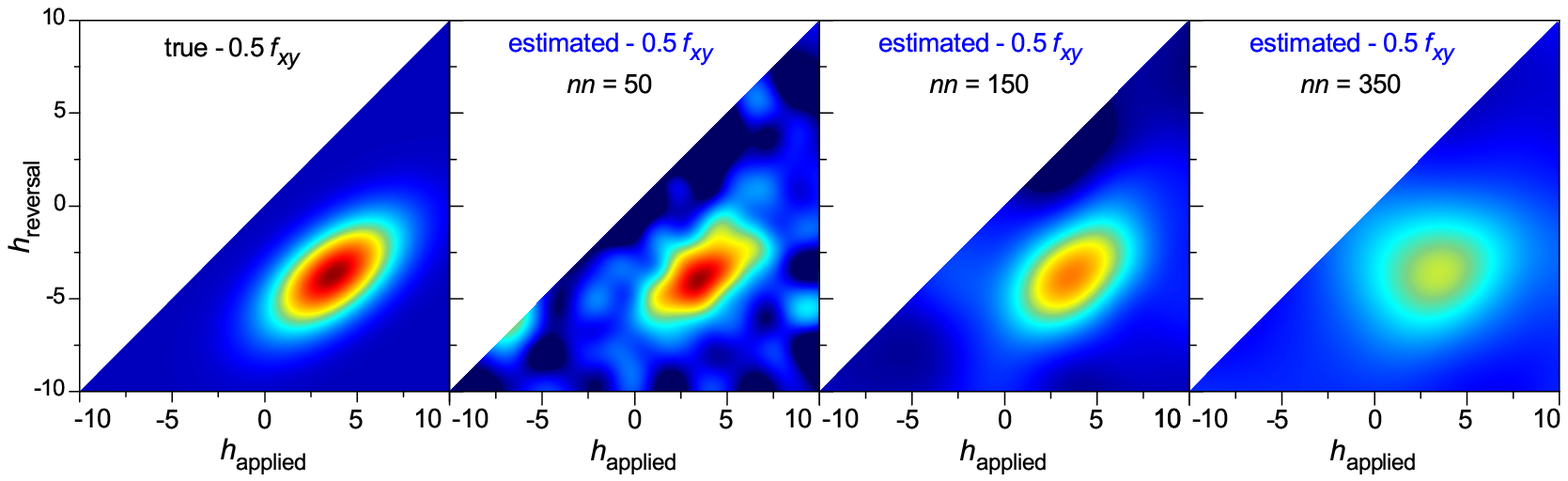}
\caption{FORC diagrams obtained with the loess method for $\protect\sigma %
\left/ R_{f}\right. =0.025$ and for different values of the smoothing
parameter $nn$, along with the true diagram.}
\label{Fig_3}
\end{figure*}

In order to study smoothing methods it is of paramount importance to use
made-up data where the true model is known. Certainly in the usual
application no representation of the true underlying function is known, but
if a method approximates a variety of functions behavior faithfully, then we
expect it to give reasonable results in other cases. Consequently, the
implemented methods were tested for accuracy on artificial data sets
constructed by adding Gaussian noise and/or outliers to different test
functions.

Our test suite consists of 15 functions, some of them with multiple features
and abrupt transitions. While all the test functions are continuous, some of
them have a derivative discontinuity at one point, what make smoothing and
the derivatives estimation quite challenging.\cite{supplementary material} A
dedicated GUI allows users to easily choose one of the functions, to add
noise, outliers, and placement of the input points. Nodes placement can be
varied from a regular grid to a complete spatial randomness. As doFORC is
mainly dedicated to FORC diagrams calculation, the test functions $%
f_{3},f_{4},f_{5},$ and $f_{6}$ provide FORCs for different types of
Preisach distributions. For all other functions the data can be generated
either in a rectangular domain or in a FORC style in the $h_{\mathrm{applied}%
}\geq h_{\mathrm{reversal}}$ half-plane. The purpose is twofold: to guide
users in the selection of an appropriate method and to provide a test suite
for assessing the accuracy of other FORC type (or two dimensional smoothing)
software tools that exist in the scientific literature.

In our tests one of the ways to characterize the error was taken to be $%
\mathrm{SSE}/\mathrm{SSM}$, where $\mathrm{SSE}=\sum\nolimits_{i=1}^{n}(\hat{%
f}_{i}-\mu _{i})^{2}$ is the sum of squared errors (deviations) from the
true test function values, and $\mathrm{SSM}=\sum\nolimits_{i=1}^{n}(\hat{f}%
_{i}-\hat{f}_{\mathrm{mean}})^{2}$ is the sum of squares of deviations about
the mean of the estimated values of dependent variable. This mean value is
the \textquotedblleft worst\textquotedblright\ regression model that always
predicts the same value irrespective of the value of the independent
variables. The coefficient of determination is defined as $r^{2}=1-\mathrm{%
SSE}/\mathrm{SSM}$. Basically $r^{2}=1$ implies no error in the estimated
values, a \textquotedblleft perfect\textquotedblright\ fit of the true data,
while $r^{2}=0$ does not implies a total lack of fit, but rather an underfit
(high bias) as will be noticed below. For brevity we will use the notations $%
h_{\mathrm{applied}},h_{\mathrm{reversal}},m$ or $x,y,f$ interchangeably.

We first present the results obtained using the test function $f_{3}$ with $%
\mu _{1}=3.7$, $\mu _{2}=0$, $\sigma _{1}=1$, $\sigma _{2}=2$, $\rho =0.2$
parameters on the $\left[ -10,10\right] \times \left[ -10,10\right] $
domain. The nodes placement was obtained starting from a $41\times 41$
regular grid by adding a uniform noise with the $0.1$ amplitude. To the true
values of the regression function evaluated at input points $\left(
x_{i},y_{i}\right) $\ we have added a normal noise with mean $0$ and
standard deviations $\sigma \left/ R_{f}\right. =0.005,0.025,0.05,0.1$,
where $R_{f}$ is the range of the test function over the given domain, the
coefficient $r^{2}$ as well as the other quantities of interest being
calculated as averages on 100 realizations of the input data (i.e., 100
different values of the seed that initialize the pseudorandom number
generator). The same results were obtained by doubling the number of
realizations. One of the realizations for each $\sigma \left/ R_{f}\right. $
value is presented in Fig. \ref{Fig_1}, where we can observe that the input
data with $\sigma \left/ R_{f}\right. =0.1$ are quite noisy. Superimposed we
present the corresponding smoothed (fitted) values in the input points,
using the four regression procedures: loess, qshep (modified quadratic
polynomial Shepard method), cshep (cubic polynomial), and tshep (cosine
series), using automatic smoothing parameter selection with the $\mathrm{AIC}%
_{\mathrm{C}}$ criterion. We observe that the noise is greatly reduced by
smoothing and the values estimated by the four methods almost overlap with
the true values, differences being mainly obtained for the most noisy data,
especially in the vicinity of the positive saturation, where the true FORCs
almost coincide. These errors give rise to unreal negative zones in the FORC
diagrams, close to the $h_{\mathrm{applied}}=h_{\mathrm{reversal}}$ diagonal
(see Fig. \ref{Fig_3}).

In Fig. \ref{Fig_2} we present the dependence of the coefficient of
determination $r^{2}$ on the number $nn$ of nearest neighbor data points
used for the local nonparametric regression, both for function and for the
first derivative $f_{x}$ and for the mixed derivative $f_{xy}$, the results
obtained for the other derivatives being similar. The symbols mark the
values given by the automatic smoothing parameter selection with the $%
\mathrm{AIC}_{\mathrm{C}}$ criterion. The $\mathrm{AIC}_{\mathrm{C}_{1}}$
and GCV criteria provide proximate values (we note that the use of $\mathrm{%
AIC}_{\mathrm{C}_{1}}$ criterion with the loess method is not recommended
for routine usage because computation time can be awful). Since in these
tests with made-up data we know the real true values, we have computed both
the Kullback-Leibler discrepancy between the smoothed estimate and the true
function, as well the true average mean squared error MSE which
characterizes the bias-variance tradeoff. All methods provide very close
values for the selected \textquotedblleft best\textquotedblright\ smoothing
parameter $nn$.

One observes a remarkable agreement between the smoothing parameters $nn_{%
\mathrm{AICC}}$ that minimize the $\mathrm{AIC}_{\mathrm{C}}$ criterion and
the maxima positions of the $r^{2}$ curves for the test function ($r^{2}@f$
). We note that for this test function the $r^{2}@f$ curves have a rather
flat maximum (and accordingly the curves $\mathrm{AIC}_{\mathrm{C}}$ have a
flat minimum), which means that different values $nn$ around the maximum can
have very similar performance on the smoothing performance. However, this is
not an abnormal situation in regression, no diagnostic technique being able
to provide an unequivocal answer, but rather a combination of diagnostic
tools together with the visual analysis of both the fitted and original data
can provide insight into the data. Hence, it is recommended to examine how
the fit varies with since sometimes fits with different smoothing parameters
can reveal features that can not be distinguished from just one
\textquotedblleft best\textquotedblright\ fit. Because both the loess and
the qshep methods use quadratic polynomials for regression, and because the
smoothed values are estimated in the input nodes, the results obtained using
these two methods are very close (the difference between them being given by
the different algorithms used for solving the least square equations).
Instead, if the smoothed values would be evaluated at points different than
the input nodes, then the loess method fit the values of $f$ on a set of
nearby nodes, while the shep methods first build nodal (associated to the
input nodes) functions that locally fits the data values and then the
estimated smoothed value is defined as a convex combination of the nodal
functions.

The $nn_{\mathrm{AICC}}$ also indicate well enough the maxima positions of
the $r^{2}@f_{x}$ and $r^{2}@f_{xy}$ curves corresponding to the first and
mixed derivatives $f_{x}$ and $f_{xy}$, respectively. This indicates that if
the local regression fits well the data within the smoothing neighborhood,
then the local slope provides a good approximation to the derivatives.
Therefore the $\mathrm{AIC}_{\mathrm{C}}$ and GCV criteria can also be used
for automatic smoothing parameter selection when the partial derivatives are
to be estimated. As expected the value of $r^{2}$ decreases when estimating
derivatives versus the function estimation, because differentiation
increases the noise due to the propagation of errors.

We notice that in general as the complexity of the local approximant
function increases the increases $r^{2}$ as well. Nevertheless, it must be
taken into account that $r^{2}$ globally characterizes the goodness of fit,
and that a more complex function can introduce oscillations into the
approximate function and so a visual analysis of the estimated data is
helpful.

In Fig. \ref{Fig_2} we have represented the dependency of $r^{2}$ on $nn$
because users usually think in terms of smoothing parameter. However, in
order to compare different amounts of smoothing being performed by different
smoothing methods the smoothers should be placed on an equal footing. Using
the same bandwidth does not give rise to comparable results as a higher
order fit is more variable but less biased. Rather, number of degrees of
freedom of a smoother is an indication of the amount of smoothing, and in
Fig. 1s\cite{supplementary material} we present the dependency of on the
degree of freedom DF1.

In Fig. \ref{Fig_3} we present the FORC diagrams obtained with the loess
method for $\sigma \left/ R_{f}\right. =0.025$ and for different values of
the smoothing parameter $nn$ along with the true diagram. Even if the input
data are rather noisy the estimated diagrams capture the main features of
the true diagram. The $nn=50$ determine an overfitted (high variance)
result, while $nn=350$ (value for which $r^{2}@f_{xy}=0$) determine an
underfitted (high bias) result.

In order to test the robustness of the four regression procedures to
outliers we have added to $20\%$ of the input data a uniform noise with
amplitude 10 times greater than that of the Gaussian noise. Fig. 2s from the
supplementary material\cite{supplementary material} shows that all methods
are capable to provide useful results. 
\begin{figure*}[tbp]
\includegraphics[width=180mm,keepaspectratio=true]{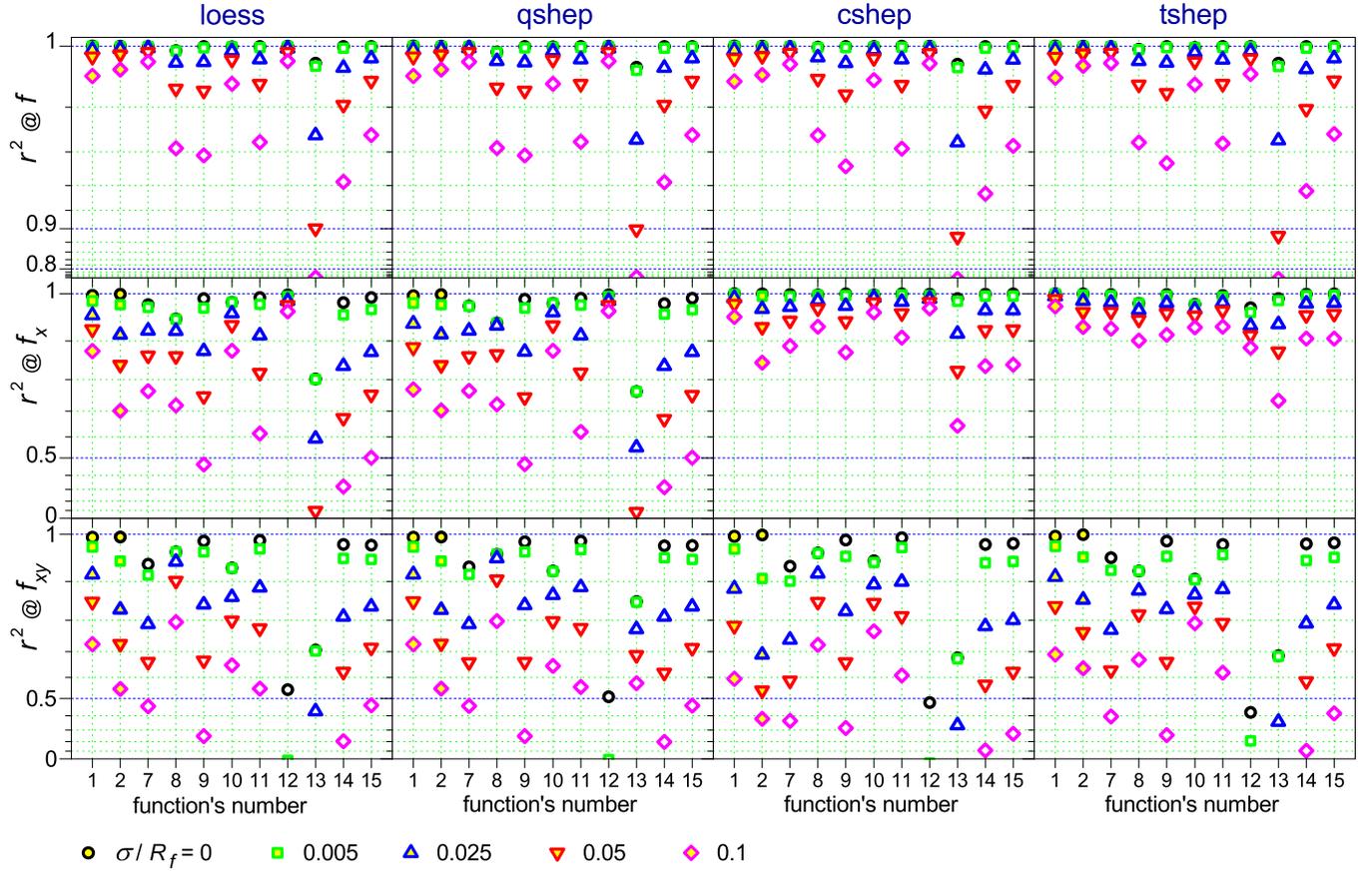}
\caption{Coefficient of determination $r^{2}$ provided by the automatic
smoothing parameter selection with the $\mathrm{AIC}_{\mathrm{C}}$ criterion
for the test functions $f_{1},f_{2},f_{7},\ldots ,f_{15}$.}
\label{Fig_4}
\end{figure*}

Figure \ref{Fig_4} summarizes some of the results of the tests performed
using functions $f_{1},f_{2}$ with $\mu _{1}=$ $\mu _{2}=0$, $\sigma _{1}=2$%
, $\sigma _{2}=4$, $\rho =0.2$, and $f_{7},\ldots ,f_{15}$ with $%
x_{c}=y_{c}=0$. The domain of the functions $f_{13},f_{14}$ was $\left[ -1,1%
\right] \times \left[ -1,1\right] $, of the function $f_{15}$ was $\left[
-3,3\right] \times \left[ -3,3\right] $, and for the other functions $\left[
-10,10\right] \times \left[ -10,10\right] $. For all functions we have used $%
41\times 41$ nodes randomly distributed throughout the corresponding
rectangular domain. Functions $f_{10},f_{12},f_{13}$ have a first derivative
discontinuity at the point $\left( 0,0\right) $. As the function $f_{13}$
has a narrow spike maximum at $\left( 0,0\right) $, covered only by 6 nodes
in our test, its $r^{2}$ values are the smallest. We note that in fact local
errors are large only near $\left( 0,0\right) $, while in the other nodes
the estimation is valuable. Although the function $f_{12}$ and its first
derivative are well estimated by all regression methods, errors are high for
the mixed derivative because the real derivative has very abrupt transitions.

It is noticed that in general the values of the first derivative are better
estimated by the regression methods with higher complexity of the local
approximant function.

\section{Conclusions}

We have presented the doFORC tool for calculating FORC diagrams of noise
scattered data. The software is also able provide both a smooth
approximation of the measured magnetization and all its partial derivatives.
The implemented nonparametric regression methods were tested for accuracy on
artificial data sets constructed by adding Gaussian noise and/or outliers to
a set of 15 test functions, some of them with multiple features and abrupt
transitions. Test performed by us have shown that all the implemented
automatic smoothing parameter selection criteria provide reliable
estimations of both smoothed magnetization and all its partial derivatives.


\begin{thebibliography}{99}
\bibitem{Mayergoyz JAP 1985} I. D. Mayergoyz, \href{http://dx.doi.org/10.1063/1.334925}%
{J. Appl. Phys.} \textbf{57}, 3803 (1985)

\bibitem{Pike JAP 1999} C. R. Pike, A. P. Roberts, and K. L. Verosub, \href{http://dx.doi.org/10.1063/1.370176}%
{J. Appl. Phys.} \textbf{85}, 6660 (1999).

\bibitem{Stancu APL 2003} A. Stancu, D. Ricinschi, L. Mitoseriu, P.
Postolache, and M. Okuyama, \href{http://dx.doi.org/10.1063/1.1623937}{Appl.
Phys. Lett.} \textbf{83}, 3767 (2003).

\bibitem{Roberts RG 2014} A. P. Roberts, D. Heslop, X. Zhao, and C.R. Pike, 
\href{http://dx.doi.org/10.1002/2014RG000462}{Rev. Geophys.} \textbf{52},
557 (2014).

\bibitem{Matau JAS 2013} F. Matau, V. Nica, P. Postolache, I. Ursachi, V.
Cotiuga, and A. Stancu, \href{http://dx.doi.org/10.1016/j.jas.2012.08.021}{%
J. Archaeol. Sci.} \textbf{40}, 914 (2013).

\bibitem{Enachescu PB 2004} C. Enachescu, R. Tanasa, A. Stancu, E. Codjovi,
J. Linares, and F. Varret, \href{http://dx.doi.org/10.1016/j.physb.2003.08.050}%
{Physica B} \textbf{343}, 15 (2004).

\bibitem{Tanasa PRB 2005} R. Tanasa, C. Enachescu, A. Stancu, J. Linares, E.
Codjovi, F. Varret, and J. Haasnoot, \href{http://dx.doi.org/10.1103/PhysRevB.71.014431}%
{Phys. Rev. B} \textbf{71}, 014431 (2005).

\bibitem{Enachescu PRB 2005} C. Enachescu, R. Tanasa, A. Stancu, F. Varret,
J. Linares, and E. Codjovi, \href{http://dx.doi.org/10.1103/PhysRevB.72.054413}%
{Phys. Rev. B} \textbf{72}, 054413 (2005).

\bibitem{Preisach 1935} F. Preisach, Z. Phys. \textbf{94}, 277 (1935).

\bibitem{Katzgraber PB 2004} H. G. Katzgraber, G. Friedman, G.T. Zim\'{a}%
nyi, \href{http://dx.doi.org/10.1016/j.physb.2003.08.051}{Physica B} \textbf{%
343}, 10 (2004).

\bibitem{Stancu JAP 2003} A. Stancu, C. Pike, L. Stoleriu, P. Postolache,
and D. Cimpoesu, \href{http://dx.doi.org/10.1063/1.1557656}{J. Appl. Phys.} 
\textbf{93}, 6620 (2003).

\bibitem{Dobrota JAP 2013} C. I. Dobrota and A. Stancu, \href{http://dx.doi.org/10.1063/1.4789613}%
{J. Appl. Phys.} \textbf{113}, 043928 (2013).

\bibitem{Almasi-Kashi PB 2014} M. Almasi-Kashi, A. Ramazani, and M.
Amiri-Dooreh, \href{http://dx.doi.org/10.1016/j.physb.2014.05.062}{Physica B}
\textbf{452}, 124 (2014).

\bibitem{Dumas PRB 2014} R. K. Dumas, P. K. Greene, D. A. Gilbert, L. Ye, C.
Zha, J. Akerman, and K. Liu, \href{http://dx.doi.org/10.1103/PhysRevB.90.104410}%
{Phys. Rev. B} \textbf{90}, 104410 (2014).

\bibitem{Gilbert SR 2014} D. A. Gilbert, G. T. Zim\'{a}nyi, R. K. Dumas, M.
Winklhofer, A. Gomez, N. Eibagi, J. L. Vicent, and K. Liu, \href{http://dx.doi.org/10.1038/srep04204}%
{Sci. Rep.} \textbf{4}, 4204 (2014).

\bibitem{Nica PB 2015} M. Nica and A. Stancu, \href{http://dx.doi.org/10.1016/j.physb.2015.07.001}%
{Physica B} \textbf{475}, 73 (2015).

\bibitem{Cimpoesu JAP 2016} D. Cimpoesu, I. Dumitru, and A. Stancu \href{http://dx.doi.org/10.1063/1.4966608}%
{J. Appl. Phys.} \textbf{120}, 173902 (2016).

\bibitem{Grafe PRB 2016} J. Gr{\"{a}}fe, M. Weigand, N. Tr{\"{a}}ger, G. Sch{%
\"{u}}tz, E. J. Goering, M. Skripnik, U. Nowak, F. Haering, P. Ziemann, and
U. Wiedwald, \href{http://dx.doi.org/10.1103/PhysRevB.93.104421}{Phys. Rev. B%
} \textbf{93}, 104421 (2016).

\bibitem{Rivas APL 2015} M. Rivas, J. C. Mart{\'{\i}}nez-Garc{\'{\i}}a, I. 
\v{S}korv\'{a}nek, J. Marcin, P. \v{S}vec, and P. Gorria, \href{http://dx.doi.org/10.1063/1.4932066}%
{Appl. Phys. Lett.} \textbf{107}, 132403 (2015).

\bibitem{Ognev N 2017} A. V. Ognev, K. S. Ermakov, A. Y. Samardak, A. G.
Kozlov, E. V. Sukovatitsina, A. V. Davydenko, L. A. Chebotkevich, A. Stancu,
and A. S. Samardak, \href{http://dx.doi.org/10.1088/1361-6528/aa564e}{%
Nanotechnology} \textbf{28}, 095708 (2017).

\bibitem{Beron NRL 2016} F. Beron, A. Kaidatzis, M. F. Velo, L. C. Arzuza,
E. M. Palmero, R. P. Del Real, D. Niarchos, K. R. Pirota, and J. M.
Garcia-Martin, \href{http://dx.doi.org/10.1186/s11671-016-1302-3}{Nanoscale
Res. Lett.} \textbf{11}, 86 (2016).

\bibitem{Markou JMMM 2013} A. Markou, K. G. Beltsios, L. N. Gergidis, I.
Panagiotopoulos, T. Bakas, K. Ellinas, A. Tserepi, L. Stoleriu, R. Tanasa,
and A. Stancu, \href{http://dx.doi.org/10.1016/j.jmmm.2013.06.009}{J. Magn. Magn. Mater.} \textbf{344}%
, 224 (2013).

\bibitem{Davies APL 2013} J. E. Davies, D. A. Gilbert, S. M. Mohseni, R. K.
Dumas, J. Akerman, and K. Liu, \href{http://dx.doi.org/10.1063/1.4813393}{%
Appl. Phys. Lett.} \textbf{103}, 022409 (2013).

\bibitem{Beron APL 2013} F. Beron, M. A. Novak, M. G. F. Vaz, G. P. Guedes,
M. Knobel, A. Caldeira, and K. R. Pirota, \href{http://dx.doi.org/10.1063/1.4816131}%
{Appl. Phys. Lett.} \textbf{103}, 052407 (2013).

\bibitem{Heslop JMMM 2005} D. Heslop and A. R. Muxworthy, \href{http://dx.doi.org/10.1016/j.jmmm.2004.09.002}%
{J. Magn. Magn. Mater.} \textbf{288}, 155 (2005).

\bibitem{Heslop GGG 2012} D. Heslop and A. P. Roberts, \href{http://dx.doi.org/10.1029/2012GC004115}%
{Geochem. Geophys. Geosyst.} \textbf{13}, Q05016 (2012).

\bibitem{FORCit 2007} G. Acton, A. Roth, and K. L. Verosub, \href{http://dx.doi.org/10. 1002/2014EO51}%
{Eos, Transactions, American Geophysical Union} \textbf{88}, 230 (2007).

\bibitem{FORCit} See \href{http://paleomag.ucdavis.edu/software-forcit.html}{%
http://paleomag.ucdavis.edu/software-forcit.html} for the FORCit software
package, which is a combination of Unix shell scripts, FORTRAN 77
subroutines, and Generic Mapping Tool (GMT) commands, all of which are
available free of charge.

\bibitem{FORCinel 2008} R. J. Harrison and J. M. Feinberg, \href{http://dx.doi.org/10.1029/2008GC001987}%
{Geochem. Geophys. Geosyst.} \textbf{9}, Q05016 (2008).

\bibitem{FORCinel} See \href{http://wserv4.esc.cam.ac.uk/nanopaleomag/?page_id=31}%
{http://wserv4.esc.cam.ac.uk/nanopaleomag/?page\textunderscore id=31} and 
\href{http://earthref.org/FORCinel}{http://earthref.org/FORCinel} for the
FORCinel software package for the calculation of FORC diagrams. The FORCinel
software is free, but runs inside Igor Pro by Wavemetrics (\href{www.wavemetrics.com}%
{www.wavemetrics.com}), which is a commercial software.

\bibitem{Cleveland 1979} W. S. Cleveland, \href{http://dx.doi.org/10.2307/2286407}%
{J. Am. Stat. Assoc.} \textbf{74}, 829 (1979).

\bibitem{Cleveland 1988} W. S. Cleveland and S. J. Devlin, \href{http://dx.doi.org/10.2307/2289282}%
{J. Am. Stat. Assoc.} \textbf{83}, 596 (1988).

\bibitem{Cleveland 1991} W. S. Cleveland and E. Grosse, \href{http://dx.doi.org/10.1007/BF01890836}%
{Stat. Comput.} \textbf{1}, 47 (1991).

\bibitem{Loader 1999} C. Loader, \textit{Local Regression and Likelihood},
Springer-Verlag New York Inc. (1999).

\bibitem{LOESS} See \href{http://www.netlib.org/a}{www.netlib.org/a} for the
FORTRAN 77 implementation of the LOESS (LOcal regrESSion) method.

\bibitem{Egli 2013} R. Egli, \href{http://dx.doi.org/}{Global Planet. Change}
\textbf{110}, 302 (2013).

\bibitem{xFORC} See \href{http://sites.google.com/site/irregularforc}{%
http://sites.google.com/site/irregularforc} for the xFORC software package
for the calculation of FORC diagrams. xFORC is freely available as an
executable file which requires Labview run-time engine.

\bibitem{Abugri JAP 2018} J. B. Abugri, P. B. Visscher, S. Gupta, P. J.
Chen, and R. D. Shull, \href{http://dx.doi.org/10.1063/1.5031786}{J. Appl.
Phys.} \textbf{124}, 043901 (2018).

\bibitem{doFORC} See \href{http://stoner.phys.uaic.ro/doFORC}{%
http://stoner.phys.uaic.ro/doFORC} for the doFORC software package for the
calculation of FORC diagrams. The doFORC is a free, portable (standalone)
application working on various operating systems.

\bibitem{Shepard 1968} D. Shepard, \textit{A two-dimensional interpolation
function for irregularly-spaced data}, \href{http://dx.doi.org/10.1145/800186.810616}%
{Proceedings of the 1968 23rd ACM national conference}, 517-524 (1968).

\bibitem{Franke-Nielson 1980} R. Franke and G. Nielson, \href{http://dx.doi.org/10.1002/nme.1620151110}%
{Internat. J. Numer. Methods Engrg.} \textbf{15}, 1691 (1980).

\bibitem{Renka 1988 Q} R. J. Renka, \href{http://dx.doi.org/10.1145/45054.45055}%
{ACM Trans. Math. Softw.} \textbf{14}, 139 (1988).

\bibitem{Renka 1999 C} R. J. Renka, \href{http://dx.doi.org/10.1145/305658.305737}%
{ACM Trans. Math. Softw.} \textbf{25}, 70 (1999).

\bibitem{Renka 1999 T} R. J. Renka and R. Brown, \href{http://dx.doi.org/10.1145/305658.305754}%
{ACM Trans. Math. Softw.} \textbf{25}, 74 (1999).

\bibitem{Shepard} See \href{www.netlib.org/toms/index.html}{%
www.netlib.org/toms/index.html} for the FORTRAN 77 implementation of the
modified Shepard methods.

\bibitem{DISLIN} See \href{http://www.mps.mpg.de/dislin}{%
http://www.mps.mpg.de/dislin} for the scientific data plotting
software/library DISLIN (Device-Independent Software LINdau). DISLIN is free
for non-commercial use.

\bibitem{GCV: Craven 1979} P. Craven and G. Wahba, \href{http://dx.doi.org/}{%
Numer. Math.} \textbf{31}, 377 (1979).

\bibitem{AIC: Akaike 1973} H. Akaike, \textit{Information theory and an
extension of the maximum likelihood principle}, \href{http://dx.doi.org/}{%
Proceedings of the "Second International Symposium on Information Theory,
1971"}, Budapest Akademiai Kiado, pp. 267-281 (1973).

\bibitem{AIC: Akaike 1974} H. Akaike, \href{http://dx.doi.org/10.1109/TAC.1974.1100705}%
{IEEE T. Automat. Contr.} \textbf{19}, 716 (1974).

\bibitem{AIC: Akaike} Akaike information criterion was developed by Hirotugu
Akaike, originally under the name \textquotedblleft An Information
Criterion. \textquotedblright It was first announced at a 1971 symposium,
whose proceedings were published in 1973. The 1973 publication was only an
informal presentation of the concepts, the formal publication being in 1974.
In this paper is shown that \textquotedblleft IC stands for information
criterion and A is added so that similar statistics, BIC, DIC etc., may
follow.\textquotedblright The 1974 paper is in the top 100 most cited
research papers of all time according to Web of Science.

\bibitem{Schwarz 1978} G. Schwarz, \href{http://www.jstor.org/stable/2958889}%
{Ann. Stat.} \textbf{6}, 461 (1978).

\bibitem{Kullback-Leibler 1951} S. Kullback and R. A. Leibler, \href{http://dx.doi.org/10.1214/aoms/1177729694}%
{Ann. Math. Stat.} \textbf{22}, 79 (1951).

\bibitem{AICC: Hurvich 1998} C. M. Hurvich, J. S. Simonoff, and C. L. Tsai, 
\href{http://dx.doi.org/10.1111/1467-9868.00125}{J. R. Statist. Soc. B} 
\textbf{60}, 271 (1998).

\bibitem{DF: Hastie 1990} T. J. Hastie and R. J. Tibshirani, \textit{%
Generalized Additive Models}, New York: Chapman \& Hall (1990).

\bibitem{supplementary material} See supplementary material at \href{...}{...%
} for the kernel and test functions used by doFORC\ software, for a
counterpart of Fig. 2 but as a function on the degree of freedom DF1, and
for a counterpart of Fig. 2 but with outliers superimposed on the Gaussian
noise.
\end{thebibliography}
\end{document}

\title{Supplementary material \smallskip \\ \hrulefill \\  \bigskip
doFORC tool for calculating first-order reversal
curve diagrams of noisy scattered data}
\author{Dorin Cimpoesu}
\author{Ioan Dumitru}
\author{Alexandru Stancu}
\affiliation{Department of Physics, Alexandru Ioan Cuza University of Iasi, Iasi 700506,
Romania}
\maketitle

\begin{longtable}[c]{|c|l|l|m{33mm}|}
\caption{Kernel functions.} \label{tab_1} \\

 \endfirsthead

 \multicolumn{4}{c}{\tablename\ \thetable{} -- continued from previous page}\\

 \hline
 \endhead
 
\multicolumn{4}{r}{{Continued on next page}} \\  \hline
\endfoot
 
 \multicolumn{4}{c}{End of Table} \\
 \hline\hline
 \endlastfoot

\hline
1 & uniform (rectangular window) & $1$ &  \includegraphics[width=33mm,keepaspectratio=true]{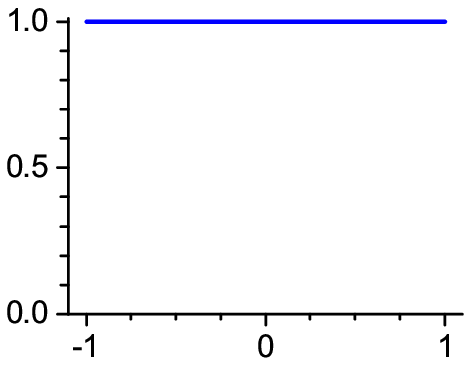} \\ \hline
2 & triangular & $1-\left\vert u\right\vert $ & \includegraphics[width=33mm,keepaspectratio=true]{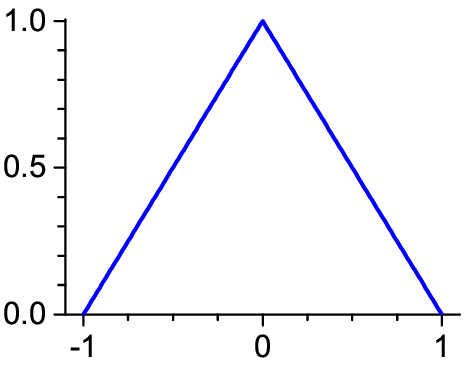} \\ \hline
3 & Epanechnikov (quadratic, parabolic) & $1-u^{2}$ & \includegraphics[width=33mm,keepaspectratio=true]{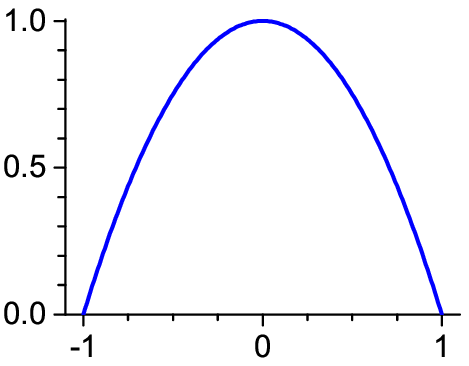} \\ \hline
4 & quartic (biweight, bisquare)& $\left( 1-u^{2}\right) ^{2}$ & \includegraphics[width=33mm,keepaspectratio=true]{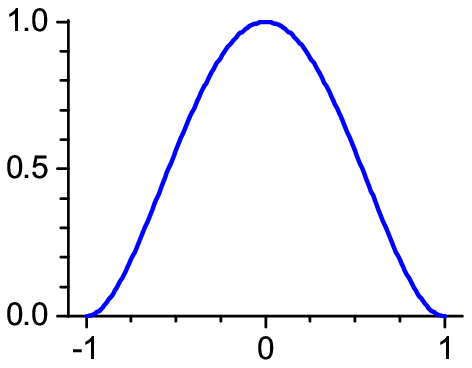} \\ \hline
5 & triweight & $\left( 1-u^{2}\right) ^{3}$ & \includegraphics[width=33mm,keepaspectratio=true]{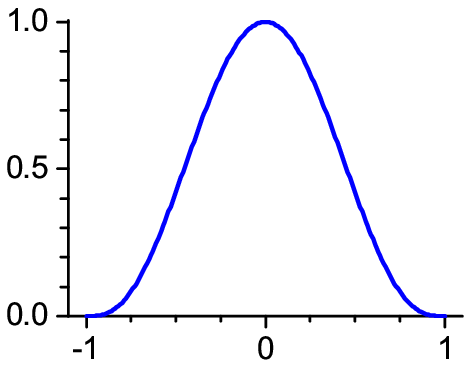} \\ \hline
6 & tricube & $\left( 1-\left\vert u\right\vert ^{3}\right) ^{3}$ & \includegraphics[width=33mm,keepaspectratio=true]{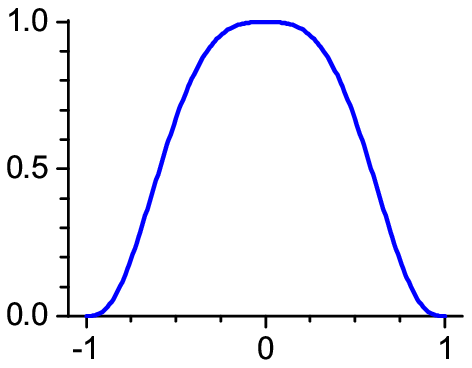} \\ \hline
7 & raised cosine (Tukey-Hanning) & $\dfrac{1+\cos \left( \pi u\right) }{2}$ & \includegraphics[width=33mm,keepaspectratio=true]{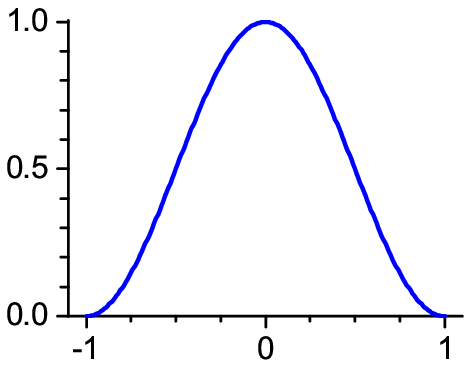}
\\ \hline
8 & cosine & $\cos \left( \dfrac{\pi }{2}u\right) $ & \includegraphics[width=33mm,keepaspectratio=true]{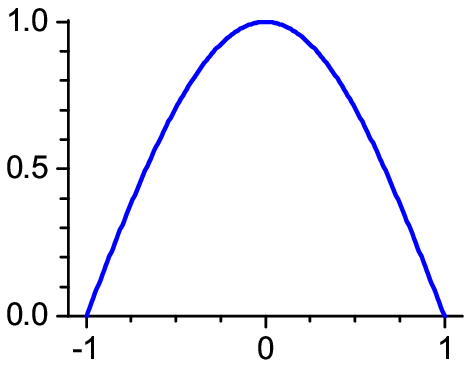} \\ \hline
9 & Gaussian & $\exp \left( -\dfrac{1}{2}\dfrac{u^{2}}{\sigma ^{2}}\right) $ & \includegraphics[width=33mm,keepaspectratio=true]{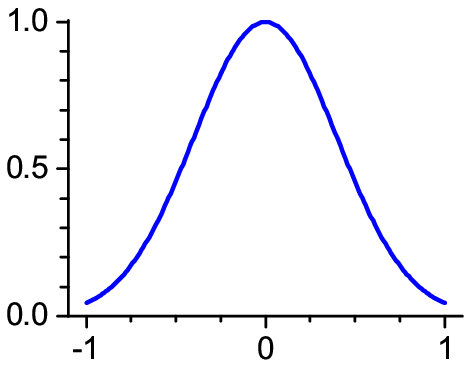}
\\ \hline
10 & exponential & $\exp \left( -\lambda \left\vert u\right\vert \right) $ & \includegraphics[width=33mm,keepaspectratio=true]{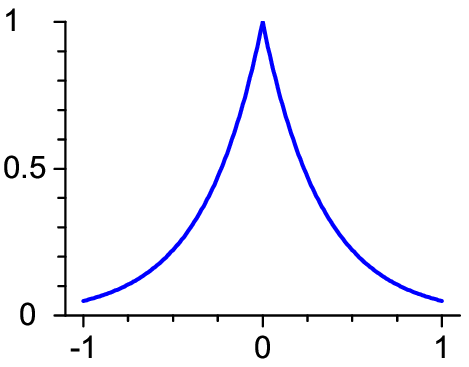}
\\ \hline
11 & inverse distance & $\dfrac{1}{1+\left\vert u\right\vert }$ & \includegraphics[width=33mm,keepaspectratio=true]{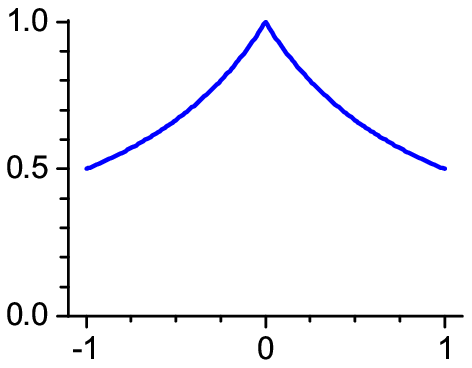} \\ \hline
12 & Cauchy & $\dfrac{1}{1+u^{2}}$ & \includegraphics[width=33mm,keepaspectratio=true]{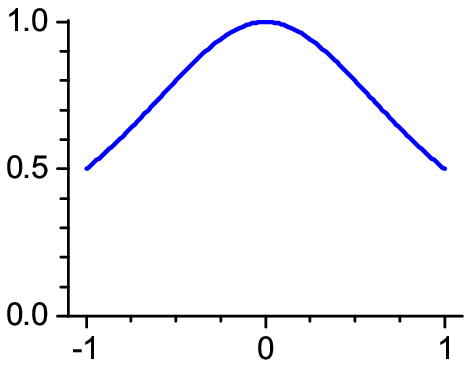} \\ \hline
13 & Parzen & $\left\{ 
\begin{array}{ll}
1-6u^{2}+6\left\vert u\right\vert ^{3} & \mathrm{,if\enspace }0\leq \left\vert
u\right\vert <0.5\;\;\; \\ 
2\left( \,1-\left\vert u\right\vert \right) ^{3} & \mathrm{,if\enspace }0.5\leq
\left\vert u\right\vert \leq 1%
\end{array}%
\right. $ & \includegraphics[width=33mm,keepaspectratio=true]{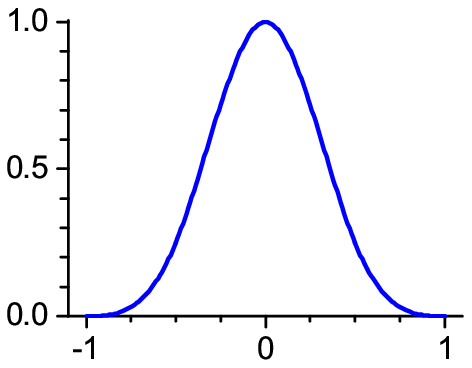} \\ \hline
14 & McLain & $\dfrac{1}{\left( \varepsilon +\left\vert u\right\vert \right)
^{2}}$ & \includegraphics[width=33mm,keepaspectratio=true]{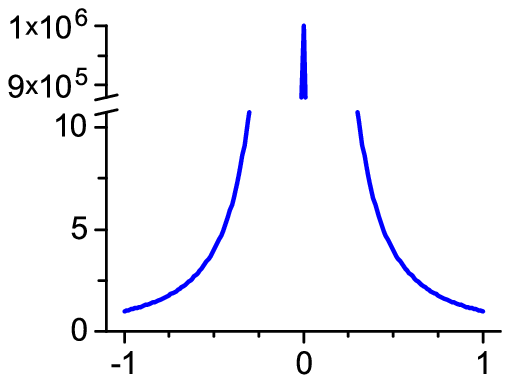} \\ \hline
15 & Franke-Nielson & $\dfrac{1-\left\vert u\right\vert }{\varepsilon
+\left\vert u\right\vert }$ & \includegraphics[width=33mm,keepaspectratio=true]{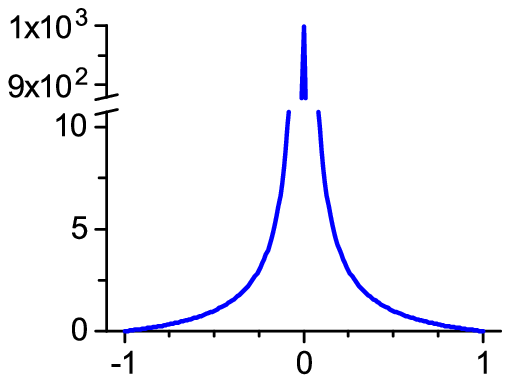} \\ \hline

\end{longtable}

\pagebreak

\begin{longtable}[c]{|c|l|}
\caption{Test functions.} \label{tab_2} \\

 \endfirsthead

 \multicolumn{2}{c}{\tablename\ \thetable{} -- continued from previous page}\\

 \hline
 \endhead
 
\multicolumn{2}{r}{{Continued on next page}} \\  \hline
\endfoot
 
 \multicolumn{2}{c}{End of Table} \\
 \hline\hline
 \endlastfoot

\hline
1 & 
\begin{tabular}{l}
Probability density function (PDF) of the bivariate normal distribution: \\ 
$%
\begin{array}{l}
{P_{\:{\mathrm{BVN}}}\left( x,y,\mu _{x},\mu _{y},\sigma _{x},\sigma
_{y},\rho \right) =} \\ 
\qquad =\dfrac{1}{2\pi \sigma _{x}\sigma _{y}\sqrt{1-\rho ^{2}}}\;\exp
\left\{ -\dfrac{1}{2\left( 1-\rho ^{2}\right) }\left[ \left( \dfrac{x-\mu
_{x}}{\sigma _{x}}\right) ^{2}+\left( \dfrac{y-\mu _{y}}{\sigma _{y}}\right)
^{2}-2\rho \left( \dfrac{x-\mu _{x}}{\sigma _{x}}\right) \left( \dfrac{y-\mu
_{y}}{\sigma _{y}}\right) \right] \right\} 
\end{array}%
$ \\ 
$\rho =\mathrm{cor}\left( {x,y}\right) =$ correlation coefficient; $%
\;-1<\rho <1$ \\ 
$\mu _{x},\;\mu _{y}=$ means \\
$\sigma _{x},\;\sigma _{y}=$ standard deviations; $\;\sigma _{x}>0;\;\sigma
_{y}>0$%
\end{tabular}
\\ \hline
2 & 
\begin{tabular}{l}
Lower cumulative distribution function (CDF) of the bivariate normal
distribution: \\ 
$CDF_{\:{\mathrm{BVN}}}\left( x,y,\mu _{x},\mu _{y},\sigma _{x},\sigma
_{y},\rho \right) =\displaystyle\int\limits_{-\infty }^{x}{d\xi
\int\limits_{-\infty }^{y}{d\eta \cdot {P_{BVN}}\left( \xi ,\eta ,\mu
_{x},\mu _{y},\sigma _{x},\sigma _{y},\rho \right) }}$%
\end{tabular}
\\ \hline
3 & 
\begin{tabular}{l}
$\mathrm{Preisach}_{\:\mathrm{BVN}}\left( x\geq y,y,\mu _{1},\mu _{2},\sigma
_{1},\sigma _{2},\rho \right) =1-\displaystyle\int\limits_{x}^{\infty }{d\xi
\;\int\limits_{y}^{\xi }{d\eta \cdot {P_{\;\mathrm{BVN}}}\left( {{\dfrac{{%
\xi -\eta }}{2}},{\dfrac{{\xi +\eta }}{2}},\mu _{1},\mu _{2},\sigma
_{1},\sigma _{2},\rho }\right) }}$ \\ 
$\mu _{1}=$ mean of the critical field distribution \\ 
$\mu _{2}=$ mean of the interaction field distribution \\ 
$\sigma _{1}=$ standard deviation of the critical field distribution \\ 
$\sigma _{2}=$ standard deviation of the interaction field distribution%
\end{tabular}
\\ \hline
4 & 
\begin{tabular}{l}
$\mathrm{Preisach}_{\:\mathrm{BVGN}}\left( x\geq y,y,\mu _{1},\mu _{2},\sigma
_{1},\sigma _{2},\rho ,\beta \right) =1-\displaystyle\int\limits_{x}^{\infty
}{d\xi \;\int\limits_{y}^{\xi }{d\eta \cdot {P_{\;\mathrm{BVGN}}}\left( {{%
\dfrac{{\xi -\eta }}{2}},{\dfrac{{\xi +\eta }}{2}},\mu _{1},\mu _{2},\sigma
_{1},\sigma _{2},\rho },\beta \right) }}$ \\ 
$%
\begin{array}{l}
{{P_{\;\mathrm{BVGN}}}\left( u,\upsilon {,\mu _{1},\mu _{2},\sigma
_{1},\sigma _{2},\rho },\beta \right) =} \\ 
\qquad =\dfrac{1}{2^{\frac{1}{\beta }}\pi \sigma _{1}\sigma _{2}\sqrt{1-\rho
^{2}}\;\Gamma \left( \dfrac{1}{\beta }\right) } \\ 
\qquad \times \exp \left\{ -\dfrac{1}{2}\left( \dfrac{1}{1-\rho ^{2}}\left[
\left( \dfrac{u-\mu _{1}}{\sigma _{1}}\right) ^{2}+\left( \dfrac{\upsilon
-\mu _{2}}{\sigma _{2}}\right) ^{2}-2\rho \left( \dfrac{u-\mu _{1}}{\sigma
_{1}}\right) \left( \dfrac{\upsilon -\mu _{2}}{\sigma _{2}}\right) \right]
\right) ^{\beta }\right\}  \\[2ex]
\qquad =\text{PDF of the bivariate generalized normal distribution}%
\end{array}%
$ \\ 
$\Gamma \left( z\right) =\text{gamma function}$ \\ 
$\beta =\text{shape parameter; }\;\beta >0$ \\ 
\colorbox{yellow}{Observation: $P_{\,\mathrm{BVN}}$ is obtained as a
particular case for $\beta =0$.}%
\end{tabular}
\\ \hline
5 & 
\begin{tabular}{l}
$%
\begin{array}{l}
{\mathrm{Preisach}_{\:\mathrm{BVSN}}\left( x\geq y,y,\mu _{1},\mu _{2},\sigma
_{1},\sigma _{2},\rho ,\delta _{11},\delta _{12},\delta _{21},\delta
_{22}\right) =} \\ 
\qquad =1-\displaystyle\int\limits_{x}^{\infty }{d\xi \;\int\limits_{y}^{\xi
}{d\eta \cdot {P_{\;\mathrm{BVSN}}}\left( {{\dfrac{{\xi -\eta }}{2}},{\dfrac{%
{\xi +\eta }}{2}},\mu _{1},\mu _{2},\sigma _{1},\sigma _{2},\rho },\delta
_{11},\delta _{12},\delta _{21},\delta _{22}\right) }}%
\end{array}%
$ \\ 
$%
\begin{array}{l}
{{{P_{\;\mathrm{BVSN}}}\left( u,\upsilon {,\mu _{1},\mu _{2},\sigma
_{1},\sigma _{2},\rho },\delta _{11},\delta _{12},\delta _{21},\delta
_{22}\right) }=} \\[1ex]
\qquad =\dfrac{1}{2\pi \sigma _{1}\sigma _{2}\sqrt{1-\rho ^{2}}\left[ \dfrac{%
1}{2}-\dfrac{1}{2\pi }\arccos \left( \widetilde{\rho }\right) \right] } \\%
[1ex]
\qquad \times \exp \left\{ -\dfrac{1}{2\left( 1-\rho ^{2}\right) }\left[
\left( \dfrac{u-\mu _{1}}{\sigma _{1}}\right) ^{2}+\left( \dfrac{\upsilon
-\mu _{2}}{\sigma _{2}}\right) ^{2}-2\rho \left( \dfrac{u-\mu _{1}}{\sigma
_{1}}\right) \left( \dfrac{\upsilon -\mu _{2}}{\sigma _{2}}\right) \right]
\right\}  \\[1ex]
\qquad \times \Phi \left[ \delta _{11}\left( u-\mu _{1}\right) +\delta
_{11}\left( \upsilon -\mu _{2}\right) \right] \,\Phi \left[ \delta
_{21}\left( u-\mu _{1}\right) +\delta _{21}\left( \upsilon -\mu _{2}\right) %
\right]  \\[3ex]
\qquad ={{{P_{\;\mathrm{BVN}}}\left( u,\upsilon {,\mu _{1},\mu _{2},\sigma
_{1},\sigma _{2},\rho }\right) }}\dfrac{\Phi \left[ \delta _{11}\left( u-\mu
_{1}\right) +\delta _{12}\left( \upsilon -\mu _{2}\right) \right] \,\Phi %
\left[ \delta _{21}\left( u-\mu _{1}\right) +\delta _{22}\left( \upsilon
-\mu _{2}\right) \right] }{\dfrac{1}{2}-\dfrac{1}{2\pi }\arccos \left( 
\widetilde{\rho }\right) } \\[2ex]
\qquad =\text{PDF of the bivariate skew normal distribution}%
\end{array}%
$ \\ 
$\delta _{11},\delta _{12},\delta _{21},\delta _{22}=\text{skewness
parameters}$ \\ 
$\Phi \left( \tau \right) =\int\limits_{-\infty }^{\tau }\dfrac{1}{\sqrt{%
2\pi }}\exp \left( -\dfrac{t^{2}}{2}\right) dt=\dfrac{1}{2}\left[ 1+\func{erf%
}\left( \dfrac{\tau }{\sqrt{2}}\right) \right] =\text{CDF of the univariate
standard normal distribution}$ \\ 
$\widetilde{\rho }=\dfrac{\delta _{21}\delta _{11}{\sigma _{1}^{2}+}\delta
_{22}\delta _{12}{\sigma _{2}^{2}+}\left( \delta _{12}\delta _{21}+\delta
_{22}\delta _{11}\right) \sigma _{1}\sigma _{2}{\rho }}{\sqrt{\left(
1+\delta _{11}^{2}{\sigma _{1}^{2}+2}\delta _{11}\delta _{12}\sigma
_{1}\sigma _{2}{\rho +}\delta _{12}^{2}{\sigma _{2}^{2}}\right) \left(
1+\delta _{21}^{2}{\sigma _{1}^{2}+2}\delta _{21}\delta _{22}\sigma
_{1}\sigma _{2}{\rho +}\delta _{22}^{2}{\sigma _{2}^{2}}\right) }}$ \\ 
\colorbox{yellow}{Observation: $P_{\,\mathrm{BVN}}$ is obtained as a
particular case for $\delta _{11}=\delta _{12}=\delta _{21}=\delta _{22}=0$.}%
\end{tabular}
\\ \hline
6 & 
\begin{tabular}{l}
$%
\begin{array}{l}
{\mathrm{Preisach}_{:\mathrm{BVSHN}}\left( x\geq y,y,\mu _{1},\mu
_{2},\sigma _{1},\sigma _{2},\rho ,\alpha _{1},\alpha _{2}\right) =} \\ 
\qquad =1-\displaystyle\int\limits_{x}^{\infty }{d\xi \;\int\limits_{y}^{\xi
}{d\eta \cdot {P_{\;\mathrm{BVSHN}}}\left( {{\dfrac{{\xi -\eta }}{2}},{%
\dfrac{{\xi +\eta }}{2}},\mu _{1},\mu _{2},\sigma _{1},\sigma _{2},\rho }%
,\alpha _{1},\alpha _{2}\right) }}%
\end{array}%
$ \\ 
$%
\begin{array}{l}
{{{P_{\;\mathrm{BVSHN}}}\left( u,\upsilon {,\mu _{1},\mu _{2},\sigma
_{1},\sigma _{2},\rho },\alpha _{1},\alpha _{2}\right) =}} \\ 
\qquad =\dfrac{4}{\alpha _{1}\,\alpha _{2}}\dfrac{1}{2\pi \sigma _{1}\sigma
_{2}\sqrt{1-\rho ^{2}}\;}\;\exp \left\{ -\dfrac{1}{2\left( 1-\rho
^{2}\right) }\left[ \left( \dfrac{2}{\alpha _{1}}\sinh \left( \dfrac{u-\mu
_{1}}{\sigma _{1}}\right) \right) ^{2}+\left( \dfrac{2}{\alpha _{2}}\sinh
\left( \dfrac{\upsilon -\mu _{2}}{\sigma _{2}}\right) \right) ^{2}\right.
\right.  \\[2ex]
\hspace{22em}\left. \left. -2\rho \left( \dfrac{2}{\alpha _{1}}\sinh \left( 
\dfrac{u-\mu _{1}}{\sigma _{1}}\right) \right) \left( \dfrac{2}{\alpha _{2}}%
\sinh \left( \dfrac{\upsilon -\mu _{2}}{\sigma _{2}}\right) \right) \right]
\right\}  \\ 
\qquad \times \cosh \left( \dfrac{u-\mu _{1}}{\sigma _{1}}\right) \,\cosh
\left( \dfrac{\upsilon -\mu _{2}}{\sigma _{2}}\right)  \\[2ex]
\qquad =\text{PDF of the bivariate sinh-normal distribution}%
\end{array}%
$ \\ 
$\alpha \;_{i}=\text{shape parameters; }\;\alpha _{i}>0$ \\ 
\colorbox{cyan}{Observation: for $\alpha >2$ the $P_{\,\mathrm{BVSHN}}$ is
multimodal.}%
\end{tabular}
\\ \hline
7 & $f_{7}\left( x,y,x_{c},y_{c}\right) =\tanh \left( x-x_{c}\right) \tanh
\left( y-y_{c}\right) $ \\ \hline
8 & $f_{8}\left( x,y,x_{c},y_{c}\right) =\sin \left( x-x_{c}\right) \:\sin
\left( y-y_{c}\right) $ \\ \hline
9 & $f_{9}\left( x,y,x_{c},y_{c}\right) =\func{sinc}\left( x-x_{c}\right) 
\:\func{sinc}\left( y-y_{c}\right) =\dfrac{\sin \left( x-x_{c}\right) }{x-x_{c}%
}\;\dfrac{\sin \left( y-y_{c}\right) }{y-y_{c}}$ \\ \hline
10 & 
\begin{tabular}{l}
$f_{10}\left( x,y,x_{c},y_{c}\right) =\sin \left( \sqrt{\left(
x-x_{c}\right) ^{2}+\left( y-y_{c}\right) ^{2}}\;\right) \equiv \sin r$ \\ 
\colorbox{magenta}{Observation: $f_{10}$ is smooth except for a first
derivative discontinuity at the point $\left( x_{c},y_{c}\right) $.}%
\end{tabular}
\\ \hline
11 & $f_{11}\left( x,y,x_{c},y_{c}\right) =\func{sinc}\left( \sqrt{\left(
x-x_{c}\right) ^{2}+\left( y-y_{c}\right) ^{2}}\;\right) \equiv \func{sinc}r$
\\ \hline
12 & 
\begin{tabular}{l}
$f_{12}\left( x,y,x_{c},y_{c}\right) =\sqrt{\left( x-x_{c}\right)
^{2}+\left( y-y_{c}\right) ^{2}}\equiv r$ \\ 
\colorbox{magenta}{Observation: $f_{12}$ is smooth except for a first
derivative discontinuity at the point $\left( x_{c},y_{c}\right) $.}%
\end{tabular}
\\ \hline
13 & 
\begin{tabular}{l}
$f_{13}\left( x,y,x_{c},y_{c}\right) =\exp \left[ -0.2\sqrt{\left( 15\left(
x-x_{c}\right) \right) ^{2}+\left( 20\left( y-y_{c}\right) \right) ^{2}}%
\;\right] \cos \left( \sqrt{\left( 15\left( x-x_{c}\right) \right) ^{2}+\left(
20\left( y-y_{c}\right) \right) ^{2}}\;\right) $ \\ 
\colorbox{magenta}{Observation: $f_{13}$ is smooth except for a first
derivative discontinuity at the point $\left( x_{c},y_{c}\right) $.}%
\end{tabular}
\\ \hline
14 & 
\begin{tabular}{l}
$%
\begin{array}{l}
{f_{14}\left( x,y,x_{c},y_{c}\right) =\left( 5^{3}\exp \left[ -5u\right]
\exp \left[ -5\upsilon \right] \right) \left( \dfrac{1}{\left( 1+\exp \left[
-5u\right] \right) \left( 1+\exp \left[ -5\upsilon \right] \right) }\right)
^{5}} \\[1ex]
\hspace{7em}\times \left( \exp \left[ -5u\right] -\dfrac{2}{1+\exp \left[ -5u%
\right] }\right) \left( \exp \left[ -5\upsilon \right] -\dfrac{2}{1+\exp %
\left[ -5\upsilon \right] }\right) 
\end{array}%
$ \\ 
where $u=x-x_{c}$; $\upsilon =y-y_{c}$ 
\end{tabular}
\\ \hline
15 & 
\begin{tabular}{l}
$%
\begin{array}{l}
{f_{15}\left( x,y,x_{c},y_{c}\right) =3\left( u-1\right) ^{2}\exp \left[
-u^{2}-\left( \upsilon +1\right) ^{2}\right] -10\left( \dfrac{u}{5}%
-u^{3}-\upsilon ^{5}\right) \exp \left[ -u^{2}-\upsilon ^{2}\right] } \\[1ex]
\hspace{7em}-\dfrac{1}{3}\exp \left[ -\left( u+1\right) ^{2}-\upsilon ^{2}%
\right] 
\end{array}%
$ \\ 
where $u=x-x_{c}$; $\upsilon =y-y_{c}$%
\end{tabular}
\\ \hline

\end{longtable}

\pagebreak

\begin{figure*}[p]
\includegraphics[width=180mm,keepaspectratio=true]{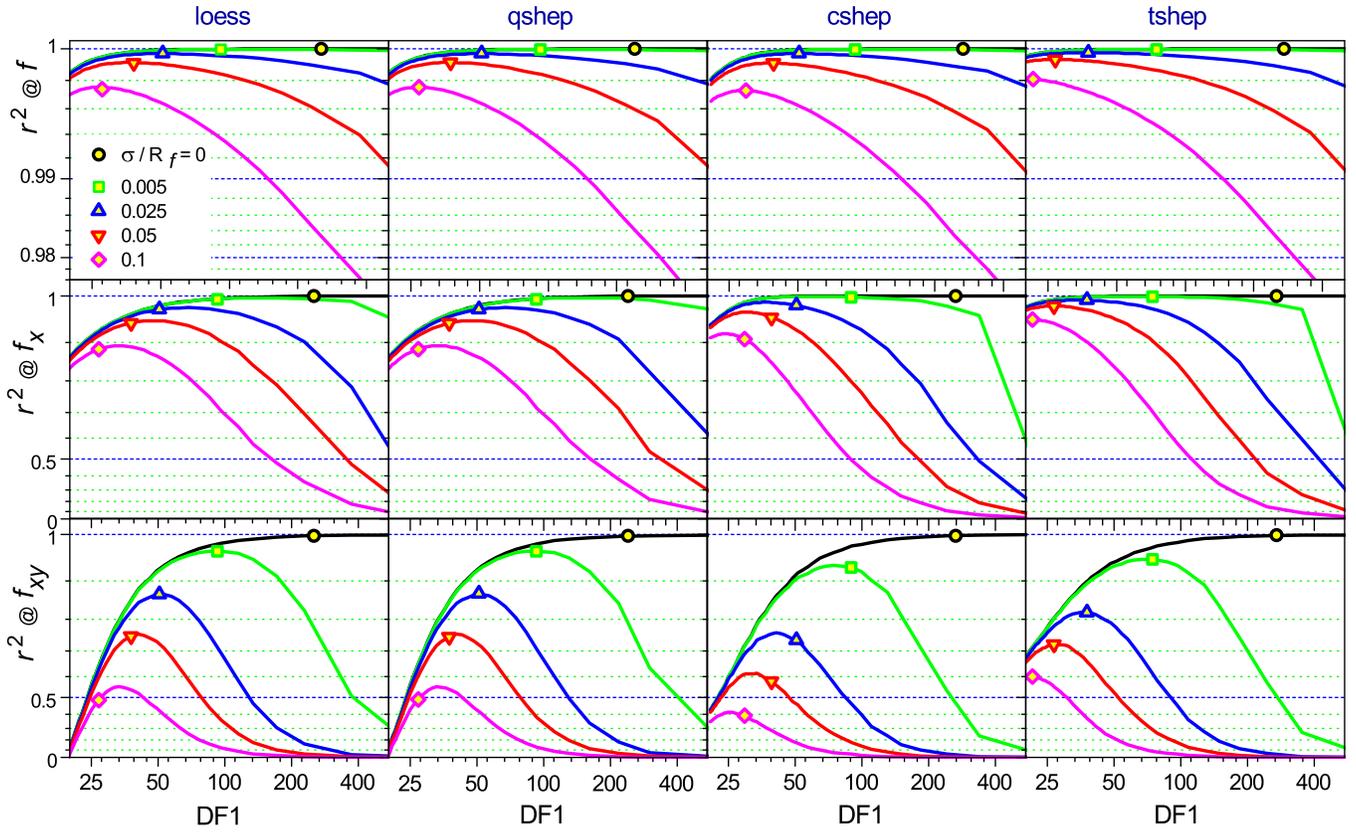}
\caption{Similar to Fig. 2, but as a function on the degree of freedom DF1.}
\label{Fig_1s}
\end{figure*}

\begin{figure*}[p]
\includegraphics[width=180mm,keepaspectratio=true]{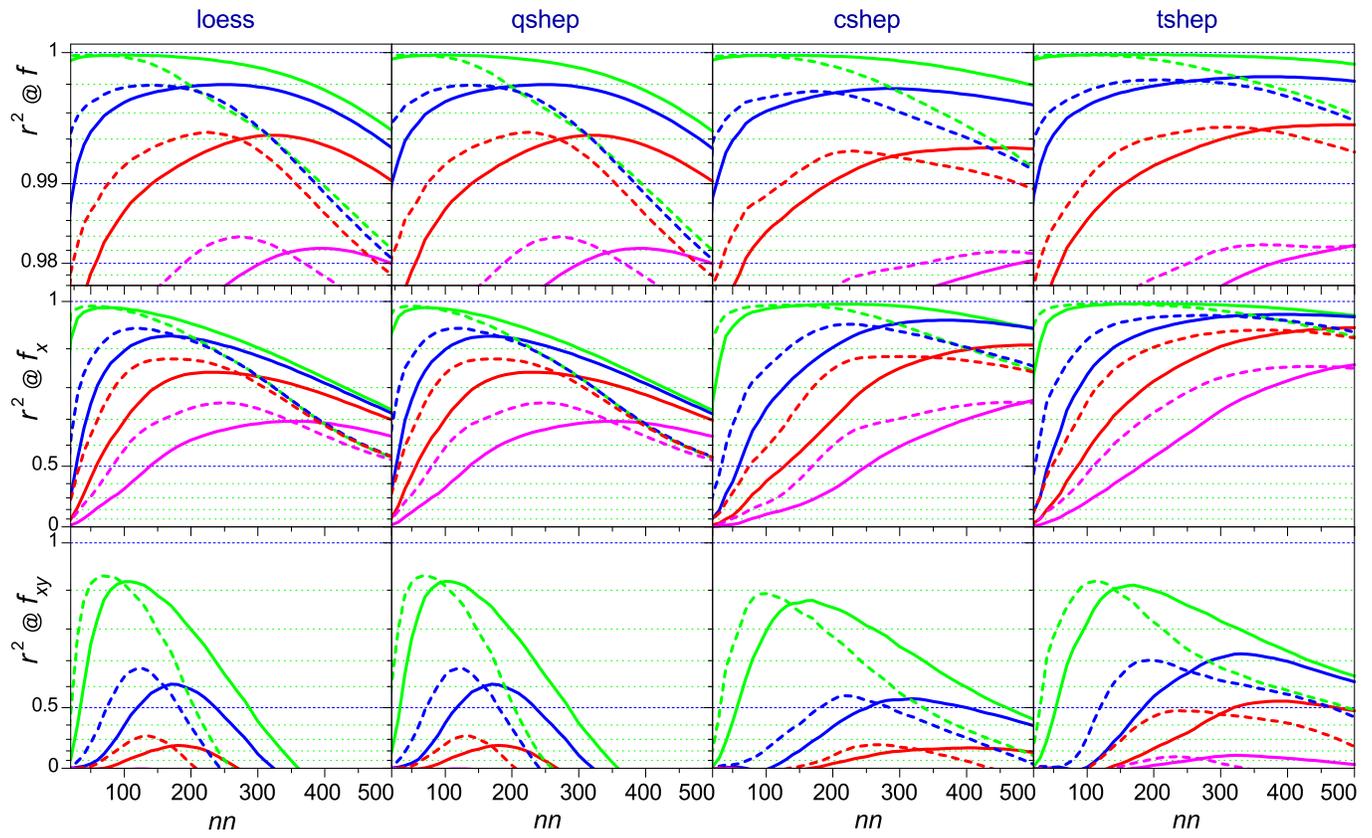}
\caption{Similar to Fig. 2, but with outliers superimposed on the Gaussian
noise. The outliers are simulated by adding to $20\%$ of the input data a
uniform noise with amplitude 10 times greater than that of the Gaussian
noise. Results obtained using a robust regression with $nnr=2$ iterations
are presented with dashed curve.}
\label{Fig_2s}
\end{figure*}